\documentclass{article}

\usepackage[preprint]{iclr2027_conference}
\usepackage{times}

\usepackage[T1]{fontenc}
\usepackage[utf8]{inputenc}
\usepackage{latexsym}
\usepackage{enumitem}
\usepackage{graphicx}
\usepackage{float}
\usepackage{xcolor}
\usepackage[most]{tcolorbox}
\tcbuselibrary{breakable,listings}
\usepackage{amsmath}
\usepackage{booktabs}
\usepackage{tabularx}
\usepackage{pifont}
\usepackage{microtype}
\usepackage{inconsolata}
\usepackage{multirow}
\usepackage[para]{footmisc}
\usepackage{hyperref}
\usepackage{subcaption}
\usepackage{url}

\usepackage{wrapfig}
\usepackage{needspace}

\graphicspath{{figure/}}
\newcommand{\cmark}{\ding{51}}
\newcommand{\xmark}{\ding{55}}
\newcommand{\rev}[1]{#1}


\title{ScholarQuest: A Taxonomy-Guided Benchmark for Agentic Academic Paper Search in Open Literature Environments}

\author{%
  Tingyue Pan, Mingyue Cheng\thanks{Corresponding author.}, Daoyu Wang, \\
  \textbf{Yitong Zhou, Jie Ouyang, Qi Liu, Enhong Chen}\\[0.3em]
  State Key Lab of Cognitive Intelligence, University of Science and Technology of China\\
  \texttt{\{pty12345,wdy030428,yitong.zhou,ouyang\_jie\}@mail.ustc.edu.cn}\\
  \texttt{\{mycheng,qiliuql,cheneh\}@ustc.edu.cn}
}

\begin{document}

\maketitle
\suppressfloats[t]
\begin{abstract}

\rev{Academic paper search is a core step in scientific research, and LLM-based search agents are emerging as a promising paradigm for iterative, intent-driven literature exploration. However, existing benchmarks offer limited coverage of research intents and insufficient support for scalable answer construction and reproducible multi-turn evaluation. We propose \textbf{ScholarQuest}, a taxonomy-guided benchmark constructed from over 1,000 computer science topics and organized around four research intents. To construct answer sets at scale, we combine multi-source retrieval, citation expansion, and multi-stage relevance adjudication. We further build \textit{\textbf{ScholarBase}}, a million-scale literature environment with unified retrieval and citation-expansion interfaces, together with a shared multi-turn search harness. We benchmark complete literature search systems and compare language models within this harness, evaluating retrieval coverage, search efficiency, and robustness across research intents. Results show that agentic methods outperform single-shot retrieval baselines, yet the best-performing search system achieves only 0.312 Recall@100 and 0.352 Recall@All. Comparisons under the shared harness reveal intent-dependent model strengths, with higher Planner token consumption not consistently yielding higher recall. We release the benchmark, retrieval backend, and evaluation harness for reproducible research\footnote{https://github.com/pty12345/ScholarQuest}.}

\end{abstract}

\section{Introduction}

\begin{wrapfigure}{r}{0.46\columnwidth}
    \vspace{-12pt}
    \centering
    \includegraphics[width=\linewidth]{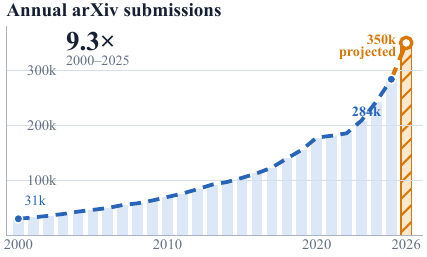}
    \vspace{-15pt}
    \caption{Annual arXiv submissions. The volume has grown dramatically year over year.}
    \label{fig:arxiv_growth}
    \vspace{-10pt}
\end{wrapfigure}

Academic paper search underpins effective knowledge discovery in scientific research \citep{timmins2005conduct, marchionini2006exploratory}. Meanwhile, the searchable literature is expanding rapidly: annual arXiv submissions\footnote{Counts are computed from the Cornell University arXiv metadata snapshot: \url{https://www.kaggle.com/datasets/Cornell-University/arxiv}.} increased 9.3-fold from 30,681 in 2000 to 284,162 in 2025 (Figure~\ref{fig:arxiv_growth}). Traditional approaches mainly rely on lexical or semantic matching, offering efficient access to large-scale literature collections \citep{shi2025spar}. However, when facing fine-grained and conditional queries, they remain limited by the single-shot ranked-list paradigm \citep{gusenbauer2020academic}. This motivates the rise of agentic paper search, where LLM-based agents can autonomously decide when and how to search, expand, and refine candidate papers based on accumulated evidence \citep{he2025pasa, pan2026paperscout}.

\begin{figure}[t]
    \centering
    \includegraphics[width=\linewidth]{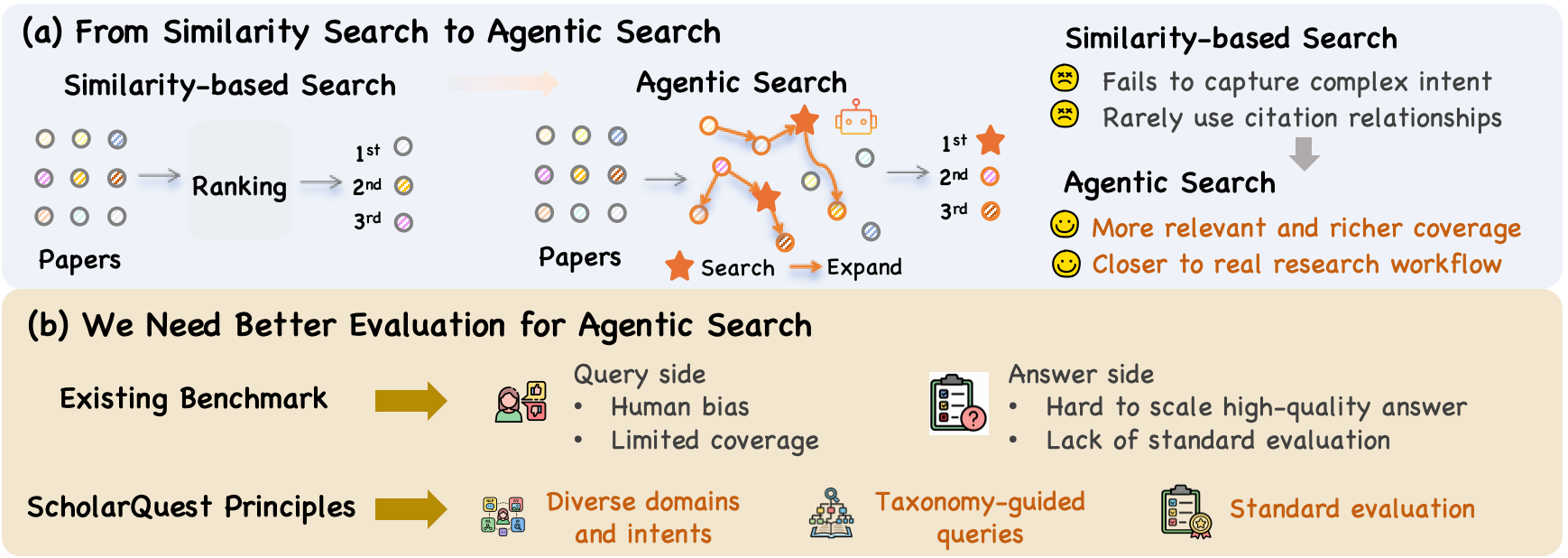}
    \caption{
    Motivation of ScholarQuest. Agentic paper search moves beyond similarity search via iterative exploration. This requires benchmarks with controlled queries and standardized evaluation.
    }
    \label{fig:motivation}
\end{figure}

\rev{Despite this progress, existing benchmarks leave three gaps in evaluating agentic academic paper search (Table~\ref{tab:benchmark_comparison}). First, manually constructed or paper-derived queries may introduce annotator bias and limit coverage of diverse research intents~\citep{he2025pasa, shi2025spar}. Second, relevant papers span different terminologies and subfields, making high-quality answer sets costly to construct at scale~\citep{he2025pasa}. Third, differences in retrieval environments and tool interfaces hinder reproducibility, while complete-system results reflect both model and workflow differences, motivating complementary model-level evaluation under a shared harness.}

\rev{To address these gaps, we propose \textbf{ScholarQuest}, a large-scale, taxonomy-guided benchmark for agentic academic paper search. Constructed from a pool of over \textbf{1,000} computer science topics, ScholarQuest covers \textbf{four} research intents: method-oriented, setting-anchored, comparison-based, and scope-controlled queries. The taxonomy provides systematic topic coverage, while the four intents enable controlled evaluation of diverse scientific search needs. To construct high-quality answer sets at scale, we develop an automated pipeline combining multi-source retrieval, citation expansion, multi-stage relevance filtering, and quality verification. By expanding beyond initial retrieval and filtering candidates in stages, the pipeline aims to improve both answer coverage and precision while reducing reliance on large-scale manual annotation. We also provide \textit{\textbf{ScholarBase}}, a million-scale literature environment based on arXiv with unified retrieval, inspection, and citation-expansion interfaces for reproducible evaluation. A shared multi-turn harness provides unified tools and controlled interaction budgets for model-level evaluation.}

\rev{We evaluate literature search systems on ScholarQuest and compare language models on ScholarQuest-100 under a shared harness, examining retrieval quality, efficiency, and robustness across research intents and answer-set sizes. Agentic systems outperform single-shot retrievers, yet the best achieves only 0.312 Recall@100 and 0.352 Recall@All, highlighting the difficulty of recovering complete relevant paper sets. Model-level comparisons reveal intent-dependent strengths, with greater Planner token consumption not consistently improving recall. We release the benchmark, retrieval backend, and harness to support reproducible research. Our contributions are summarized as follows:}




\begin{itemize}[itemsep=2pt, topsep=1pt, parsep=1.5pt]
    \item \rev{We propose \textbf{ScholarQuest}, a large-scale, taxonomy-guided benchmark for agentic academic paper search, constructed from a pool of over 1,000 CS topics and covering four research intents, with a million-scale paper retrieval backend \textit{\textbf{ScholarBase}}.}

    \item We build a scalable construction pipeline that integrates taxonomy-guided query generation, multi-source retrieval, citation expansion, relevance filtering, and quality verification.

    \item \rev{We conduct complementary system- and model-level evaluations, revealing limitations in retrieval coverage, scope control, and search efficiency, as well as intent-dependent differences in model performance and Planner token usage.}
\end{itemize}

\section{Related Work}

\subsection{Academic Literature Search}

\rev{Academic literature search underpins scientific knowledge discovery~\citep{timmins2005conduct,marchionini2006exploratory}. Sparse retrievers such as BM25 estimate lexical relevance through probabilistic term weighting, enabling efficient large-scale search but remaining vulnerable to vocabulary mismatch~\citep{robertson2009probabilistic}. Dense methods, including DPR for query--document matching and SPECTER for scientific-document representation, encode text in semantic spaces to alleviate lexical mismatch, but fixed-length representations can limit precise retrieval over long documents~\citep{karpukhin2020dense,cohan2020specter,luan2021sparse}. Hybrid methods such as RRF fuse sparse and dense rankings for more robust retrieval, yet still produce a fixed ranked list~\citep{cormack2009reciprocal}. LLM-enhanced methods further support query understanding, reformulation, reranking, and relevance filtering, but predefined workflows limit adaptive exploration~\citep{shi2025spar,he2025pasa}. These limitations motivate agentic literature search for iterative, intent-driven exploration.}

\begin{table}[t]
\centering
\fontsize{8.2pt}{10pt}\selectfont
\caption{
\rev{Comparison of literature search benchmarks. \#Queries and \#Papers count evaluation queries and globally distinct gold papers, respectively. Automation indicates construction without per-item expert curation; Intent Control indicates explicit query-intent control.}
}
\setlength{\tabcolsep}{2pt}
\renewcommand{\arraystretch}{1.08}

\resizebox{\linewidth}{!}{%
\begin{tabular}{@{}lcccccc@{}}
\toprule
\textbf{Benchmark}
& \textbf{Automation}
& \textbf{Intent Control}
& \textbf{\#Queries}
& \textbf{\#Papers}
& \textbf{Construction}
& \textbf{Topic Scope} \\
\midrule
RealScholar~\citep{he2025pasa}
& \xmark
& \xmark
& 50
& 772
& Query-first
& CV / NLP \\
AutoScholar~\citep{he2025pasa}
& \cmark
& \xmark
& 1,000
& 2,009
& Answer-first
& Seed Papers \\
SPARBench~\citep{shi2025spar}
& \xmark
& \xmark
& 50
& 523
& Query-first
& AI / NLP \\
\midrule
\textbf{ScholarQuest}
& \textbf{\cmark}
& \textbf{\cmark}
& \rev{\textbf{1,111}}
& \rev{\textbf{49,162}}
& \rev{\textbf{Query-first}}
& \rev{\textbf{1,000+ CS topics}} \\
\bottomrule
\end{tabular}
}

\label{tab:benchmark_comparison}
\end{table}

\subsection{Agentic Literature Search and Evaluation}

\rev{LLM-based agents have recently been explored for academic literature search. Agentic workflow-based approaches such as PaSa formulate literature search as a multi-step process that searches, inspects, and selects relevant papers~\citep{he2025pasa}. SPAR further integrates query understanding, multi-source retrieval, citation-based exploration, and reranking for training-free literature search~\citep{shi2025spar}. Building on this line of work, PaperScout improves tool-use autonomy by enabling the agent to decide \textit{when} and \textit{how} to invoke search and expansion actions according to the evolving search state~\citep{pan2026paperscout}. These studies demonstrate the promise of agentic exploration beyond single-shot retrieval, including for deep-research agents that require trustworthy scholarly evidence~\citep{wang2025paperarena,cheng2026mind2report,jin2025search}. Existing benchmarks, however, trade off scale, realism, and coverage: AutoScholarQuery scales through pairs derived from \textit{Related Work} sections but may inherit topic and citation biases, whereas RealScholarQuery and SPARBench use real-world or expert-screened queries yet remain small and lack systematic topic and intent coverage~\citep{he2025pasa,shi2025spar}. ScholarQuest complements them with taxonomy-grounded topics, controlled query intents, and broader answer pools for multi-dimensional agent evaluation.}

\section{ScholarQuest}
\label{sec:benchmark_construction}

In this section, we first establish a clear task formulation, then describe the benchmark construction pipeline, followed by the four research intents used to evaluate different retrieval capabilities.



\subsection{Task Formulation}

ScholarQuest evaluates academic paper search as a multi-turn retrieval task. Given a research query $q$ and a literature database $\mathcal{D}$, a search system iteratively analyzes the current paper pool $\mathcal{P}_t$ and issues retrieval or exploration requests to $\mathcal{D}$. The returned papers are merged into $\mathcal{P}_{t+1}$, allowing the system to accumulate candidate papers and update search state over multiple turns. The system returns a ranked list $\hat{\mathcal{P}}$, which is evaluated against the ground-truth relevant paper set $\mathcal{P}^{*}$.


\Needspace{6\baselineskip}
\subsection{Benchmark Construction}

\begin{wrapfigure}{R}{0.48\columnwidth}
    \centering
    \vspace{-0.28in}
    \includegraphics[width=\linewidth]{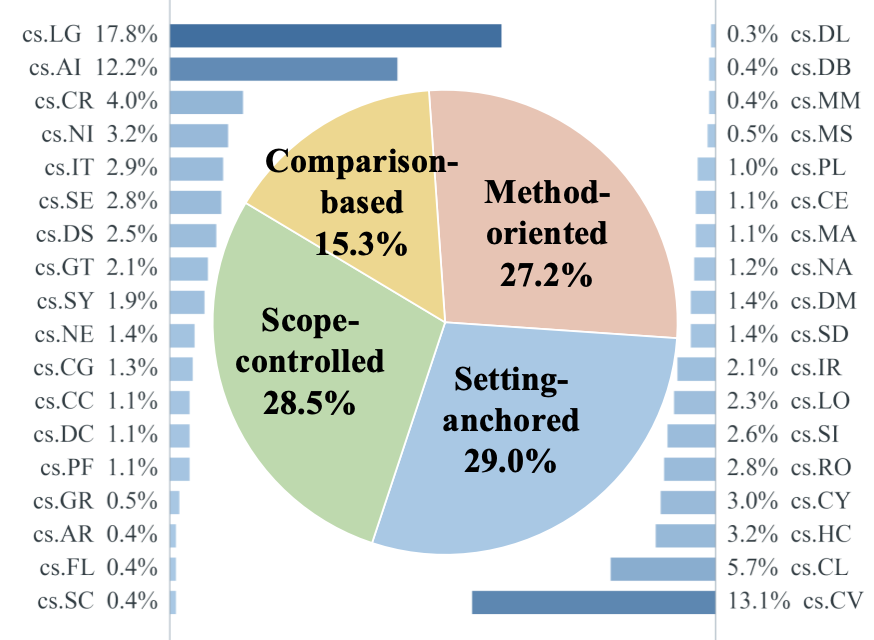}
    \vspace{-0.25in}
    \caption{Dataset istribution of ScholarQuest. Side bars show query counts across arXiv CS categories, while the center pie shows query intent distribution.}
    \label{fig:data_distribution}
    \vspace{-20pt}
\end{wrapfigure}

\subsubsection{Query Generation}
\label{sec:query_generation}

We first collect over 1,600 topics from the ACM Computing Classification System (CCS), which provides a poly-hierarchical organization of computing topics~\citep{rous2012major}. To focus the benchmark on computer science paper search, we utilize Qwen3-Max \citep{yang2025qwen3} to map each ACM CCS topic to one or more arXiv subject categories according to the arXiv category taxonomy\footnote{\url{https://arxiv.org/category_taxonomy}}. We retain topics assigned to arXiv CS categories and discard topics outside the computer science scope. \rev{This process results in a pool of over 1,000 eligible CS taxonomy nodes, which serves as the basis for frequency-weighted topic-seed sampling.}

\rev{Using topic seeds sampled from this pool, we further utilize Qwen3-Max to generate research queries under four intent types: \textbf{method-oriented}, \textbf{setting-anchored}, \textbf{comparison-based}, and \textbf{scope-controlled}. This design allows us to systematically cover diverse academic search intents rather than relying on queries derived from individual papers. Detailed sampling, validation, deduplication, and filtering procedures are provided in Appendix~\ref{app:construction_dedup}. We further detail the topic coverage of the released set and intent distributions across query selection in Appendix~\ref{app:query_validity}.}

\begin{figure}[t]
    \centering
    \includegraphics[width=0.99\linewidth]{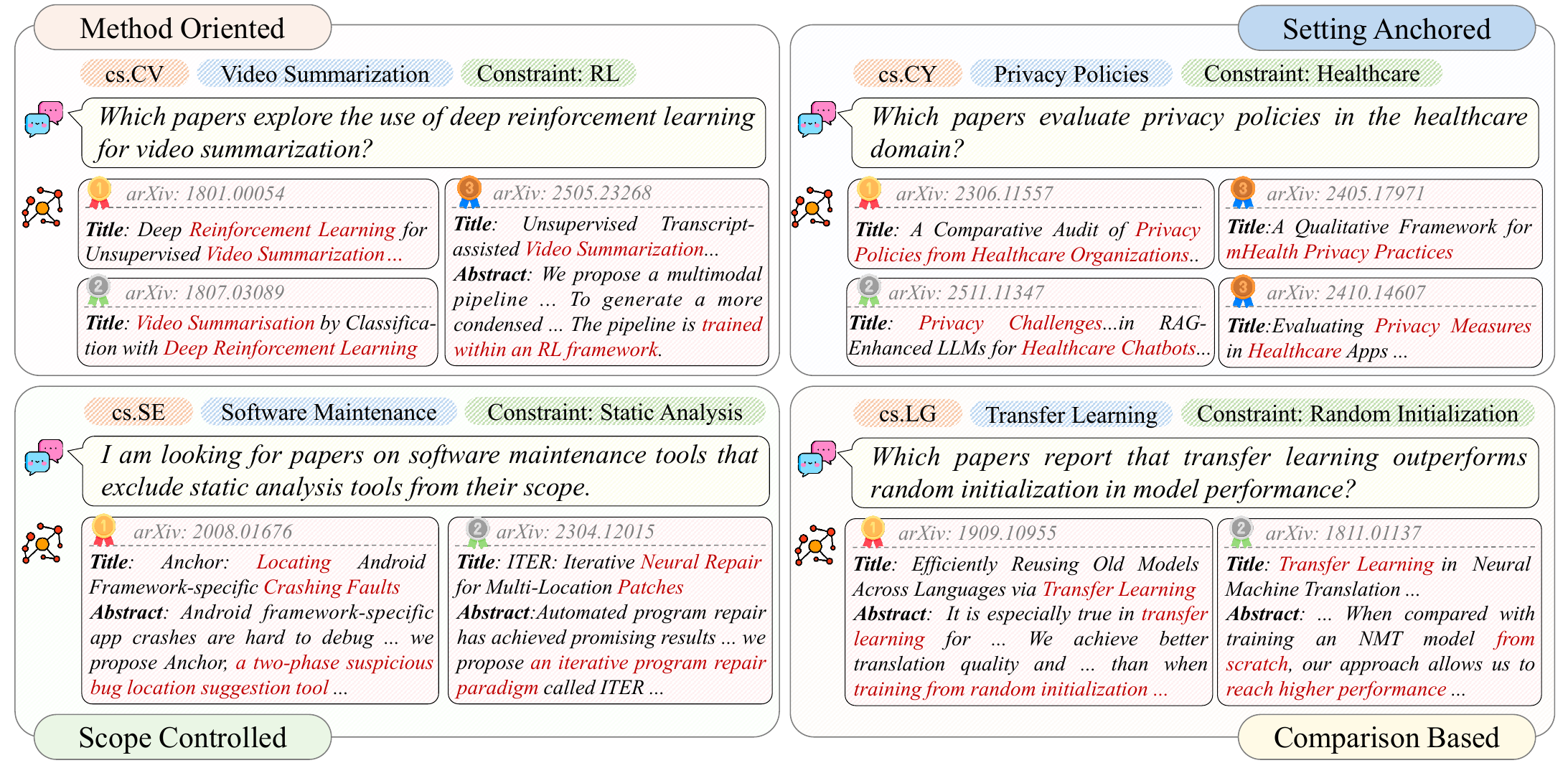}
    \caption{Cases of the four research-intent types in ScholarQuest. Each quadrant presents a query with its domains, topic seed, constraints and example gold papers from the answer set. }
    \label{fig:query_type}
\end{figure}

\subsubsection{Answer Discovery}
\label{sec:answer_discovery}

\paragraph{Multi-source Candidate Retrieval.}

In the retrieval stage, we collect a candidate set that serves as the starting point for expansion. For each query, we use the original query and LLM-generated rewrites, including method-focused, task-focused, broader-topic, narrower-topic, and terminology-based formulations. These queries are issued to multiple retrieval sources, including Google Search API\footnote{\url{https://serper.dev}}, arXiv API\footnote{\url{https://info.arxiv.org}}, and Semantic Scholar API\footnote{\url{https://www.semanticscholar.org}}. The union of their results forms the retrieval-based candidate pool $\mathcal{C}^{ret}_q$.

\paragraph{Citation-Graph Expansion.}

Since direct retrieval may miss relevant papers with different terminology, we further expand the retrieved candidates through citation relations. Starting from high-confidence papers in $\mathcal{C}^{ret}_q$, we collect their references and citing papers when available, and selectively perform second-hop expansion from high-confidence neighbors while pruning noisy branches. We merge the expanded papers with $\mathcal{C}^{ret}_q$ to form the final candidate pool $\mathcal{C}_q$. This step complements direct retrieval by recovering papers that are connected through scholarly relations but may be missed by lexical or semantic matching.

\paragraph{Multi-stage Relevance Adjudication.}

We derive the final answer set from the expanded candidate pool through multi-stage relevance adjudication. First, we remove clear mismatches using retrieval scores, metadata signals, and a small relevance model, while keeping the filtering stage recall-oriented. The remaining candidates are evaluated by multiple LLM-based relevance judges using the query, paper title, and abstract. Each judge assigns a score from 0 to 2, denoting mismatch, partial match, and strict match, respectively. After aggregating the scores into $\text{Rel}(q,p)$, we select papers above threshold $\tau=2$:
\begin{equation}
    \mathcal{A}_q =
    \{p \in \mathcal{C}_q \mid \text{Rel}(q, p) \geq \tau \}.
\end{equation}
Borderline and high-disagreement cases are further adjudicated to reduce labeling errors.

\subsubsection{Human-in-the-Loop Curation}
\label{sec:human_loop_curation}

Building on these automatically constructed answer sets, we curate a high-fidelity subset through stratified selection and iterative human review. For each selected query, Ph.D. researchers review existing answers against constraints on methods, settings, scope, or comparisons, while searching the frozen literature environment for missing papers. Based on the review findings, the Ph.D. researchers refine ambiguous queries and use the revised constraints to guide further retrieval and reassessment of affected papers, repeating this cycle until convergence to improve query clarity, label accuracy, and answer coverage. Convergence criteria and quality controls are detailed in Appendix~\ref{app:curation_convergence}.

\subsection{Key Features of ScholarQuest}

ScholarQuest covers four representative research intents in academic paper search. As shown in Figure \ref{fig:query_type}, each type reflects a common search scenario and targets a distinct retrieval capability.

\vspace{-0.01in}

\paragraph{Method-oriented Queries.}
This dimension focuses on finding papers that use, extend, or analyze a specific method. It tests whether systems can identify methodological relevance beyond surface-level keyword overlap.

\vspace{-0.01in}

\paragraph{Setting-anchored Queries.}
This category targets papers under a specific task, dataset, or experimental setting. It evaluates whether systems can satisfy fine-grained contextual constraints rather than retrieve broadly related papers.

\vspace{-0.01in}

\paragraph{Scope-controlled Queries.}
This feature specifies explicit boundaries for the desired paper set, such as a target domain, task, or model family. It evaluates whether systems can control retrieval breadth and avoid loosely relevant results.

\vspace{-0.01in}

\paragraph{Comparison-based Queries.}
This aspect focuses on papers that compare methods, assumptions, results, or technical claims. It tests whether systems can capture relational search intents involving multiple research objects.


\subsection{\rev{Evaluation Harness}}
\label{sec:evaluation_harness}

\rev{To further evaluate the literature search capabilities of foundation models, we build a standardized harness that supports multi-turn retrieval through a unified tool interface over the shared \textit{\textbf{ScholarBase}} environment. The harness exposes two atomic tools: \texttt{Search} retrieves candidate papers from the literature database using model-generated queries, whereas \texttt{Expand} augments the candidate set by retrieving the references and citing papers of a specified paper in the current pool. At each round, the foundation model selects the next tool calls and their arguments based on the research query and current paper-pool state. The harness handles candidate relevance scoring, paper-pool updates, and final result ranking, while logging per-round actions and changes to the paper pool to consistently evaluate retrieval quality and search efficiency.}

\section{Experiment}

\vspace{-0.05in}

\rev{We first evaluate complete literature search systems on ScholarQuest, then compare model-driven search quality and Planner token usage under the shared harness on ScholarQuest-100. We further analyze search efficiency, robustness, and shared failure patterns.}

\subsection{Experimental Setup}

\vspace{-0.05in}

\paragraph{Evaluation Sets.}
We use two complementary evaluation sets. \textbf{ScholarQuest} is the complete output of Sections~\ref{sec:query_generation}--\ref{sec:answer_discovery} and contains 1,111 queries, with 302/322/317/170 method/setting/scope/comparison queries (Figure~\ref{fig:data_distribution}). We use this set for the system-level evaluation. \textbf{ScholarQuest-100} contains 100 queries produced from ScholarQuest by Section~\ref{sec:human_loop_curation}, with 25/25/25/25 queries across the same intents. We use this set for the model-level evaluation under the shared harness in Section~\ref{sec:model_level_results}. Its frozen queries and graded judgments support fine-grained evaluation of ranking quality and constraint satisfaction.

\begin{table*}[!t]
\centering
\footnotesize
\setlength{\tabcolsep}{2.0pt}
\renewcommand{\arraystretch}{0.95}
\caption{Recall-oriented results on ScholarQuest. We report Recall@25, Recall@100, and Recall@All across four query types and the overall query set. Agentic entries are means over three independent runs. The \textbf{bold} indicates the best result, while \underline{underline} indicates the second-best.}
\resizebox{0.98\textwidth}{!}{
\begin{tabular}{@{}lccccccccccccccc@{}}
\toprule
\multirow{2}{*}{\textbf{Method}}
& \multicolumn{3}{c}{\textbf{Method Oriented}}
& \multicolumn{3}{c}{\textbf{Setting Anchored}}
& \multicolumn{3}{c}{\textbf{Scope Controlled}}
& \multicolumn{3}{c}{\textbf{Comparison Based}}
& \multicolumn{3}{c}{\textbf{Overall}} \\
\cmidrule(lr){2-4}
\cmidrule(lr){5-7}
\cmidrule(lr){8-10}
\cmidrule(lr){11-13}
\cmidrule(l){14-16}
& \textbf{R@25} & \textbf{R@100} & \textbf{R@All}
& \textbf{R@25} & \textbf{R@100} & \textbf{R@All}
& \textbf{R@25} & \textbf{R@100} & \textbf{R@All}
& \textbf{R@25} & \textbf{R@100} & \textbf{R@All}
& \textbf{R@25} & \textbf{R@100} & \textbf{R@All} \\
\midrule
Dense  Retrieval
& 0.155 & 0.300 & \underline{0.403}
& 0.110 & 0.219 & 0.306
& 0.029 & 0.087 & 0.159
& 0.136 & 0.234 & \underline{0.326}
& \rev{0.103} & \rev{0.206} & \rev{0.293} \\

Hybrid Retrieval
& 0.153 & 0.303 & 0.344
& 0.115 & 0.226 & 0.257
& 0.038 & 0.091 & 0.111
& 0.133 & 0.250 & 0.276
& \rev{0.106} & \rev{0.212} & \rev{0.242} \\

\midrule
Google Search
& 0.106 & 0.146 & 0.167
& 0.059 & 0.094 & 0.117
& 0.003 & 0.006 & 0.013
& 0.122 & 0.135 & 0.143
& \rev{0.065} & \rev{0.089} & \rev{0.105} \\

Google Scholar
& 0.084 & 0.111 & 0.143
& 0.073 & 0.102 & 0.120
& 0.004 & 0.010 & 0.010
& 0.107 & 0.125 & 0.130
& \rev{0.062} & \rev{0.082} & \rev{0.096} \\

Semantic Scholar
& 0.064 & 0.085 & 0.101
& 0.103 & 0.116 & 0.125
& 0.001 & 0.002 & 0.002
& 0.089 & 0.103 & 0.103
& \rev{0.061} & \rev{0.073} & \rev{0.080} \\

DeepXiv \citep{qian2026deepxiv}
& 0.091 & 0.214 & 0.341
& 0.068 & 0.181 & 0.304
& 0.007 & 0.023 & 0.064
& 0.039 & 0.118 & 0.286
& \rev{0.052} & \rev{0.135} & \rev{0.243} \\

\midrule
PaSa~\citep{he2025pasa}
& \underline{0.245} & \underline{0.345} & 0.366
& \underline{0.230} & \underline{0.296} & \underline{0.312}
& \textbf{0.103} & \textbf{0.193} & \textbf{0.242}
& \underline{0.256} & \underline{0.294} & \underline{0.324}
& \rev{\underline{0.202}} & \rev{\underline{0.280}} & \rev{\underline{0.309}} \\

SPAR~\citep{shi2025spar}
& 0.242 & 0.333 & 0.351
& 0.223 & 0.282 & 0.293
& \underline{0.102} & \underline{0.188} & 0.222
& 0.246 & 0.280 & 0.301
& \rev{0.197} & \rev{0.269} & \rev{0.290} \\

PaperScout~\citep{pan2026paperscout}
& \textbf{0.275} & \textbf{0.408} & \textbf{0.451}
& \textbf{0.246} & \textbf{0.341} & \textbf{0.366}
& 0.091 & 0.182 & \underline{0.239}
& \textbf{0.275} & \textbf{0.327} & \textbf{0.358}
& \rev{\textbf{0.214}} & \rev{\textbf{0.312}} & \rev{\textbf{0.352}} \\
\bottomrule
\end{tabular}
}

\label{tab:main_table}
\end{table*}

\vspace{-0.05in}

\begin{figure}[!t]
    \centering
    \includegraphics[width=\linewidth]{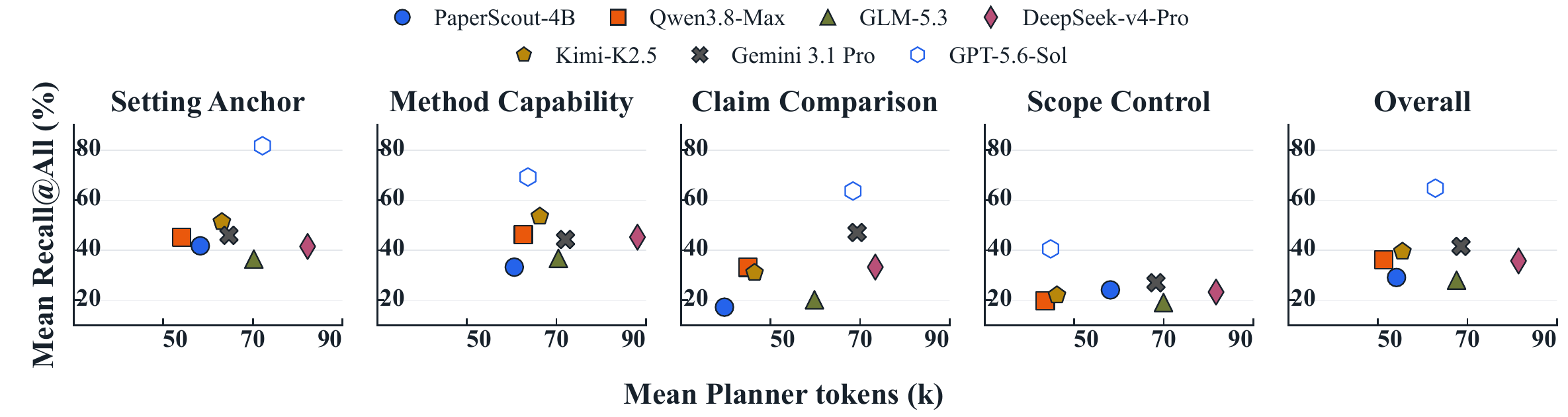}
    \caption{Mean token usage versus Recall@All on ScholarQuest-100 under the shared harness. The x-axis shows the total input and output tokens consumed by the model over a complete multi-turn trajectory, averaged across valid queries.}
    \label{fig:model_level_results}
\end{figure}

\paragraph{Environment.}
We build \textit{\textbf{ScholarBase}}, an open-source paper-search environment constructed from the S2 PaperData snapshot~\citep{lo-wang-2020-s2orc}, which retains arXiv papers with abstracts, and stores paper metadata and citation relations. It supports BM25 \citep{harman1995overview} sparse retrieval with SQLite FTS5\footnote{\url{https://sqlite.org}}, dense retrieval with Qdrant over BGE-M3 \citep{chen2024bge} title--abstract embeddings, and RRF-based hybrid retrieval \citep{cormack2009reciprocal}. The environment further provides RESTful APIs for semantic search, title matching, metadata lookup, and citation/reference traversal, enabling all methods to search, inspect, and expand papers under the same controlled literature backend.

\paragraph{Baselines.}
\rev{We evaluate both literature search systems and foundation models. At the system level, we compare three groups: standard retrieval baselines implemented in \textit{\textbf{ScholarBase}}, including dense retrieval and RRF-based hybrid search; widely used academic search systems, including Google Search\footnote{\url{https://serper.dev}}, Google Scholar\footnote{\url{https://scholar.google.com}}, Semantic Scholar API\footnote{\url{https://www.semanticscholar.org}}, and DeepXiv~\citep{qian2026deepxiv}; and open-source agentic literature search systems, including PaSa~\citep{he2025pasa}, SPAR~\citep{shi2025spar}, and PaperScout~\citep{pan2026paperscout}. At the model level, we compare Qwen3.8-Max, Kimi-K2.5, GLM-5.3, DeepSeek-v4-Pro, Gemini 3.1 Pro, and GPT-5.6-Sol using the shared harness described in Section~\ref{sec:evaluation_harness}, with PaperScout-4B included as a small-model reference.}

\vspace{-0.05in}

\paragraph{Metrics.}
On ScholarQuest, we report Recall@25, Recall@100, and Recall@All to measure answer coverage. Recall@$k$ uses the top $k$ normalized, deduplicated papers, whereas Recall@All uses the complete ranking returned at termination. Query-level macro Precision and F1 at the same cutoffs are provided in Appendix~\ref{app:complementary_evaluation}. On ScholarQuest-100, we report Recall@All for model-level coverage comparisons; graded judgments also support NDCG, MRR, and Precision/Recall@5 and @10. For process analysis, we report interaction rounds, tool calls, observed candidates, Recall@100 per 100 candidates, and mean Planner tokens. We also examine robustness across research intents and answer-set sizes.

\begin{figure}[t]
    \centering
    \includegraphics[width=\linewidth]{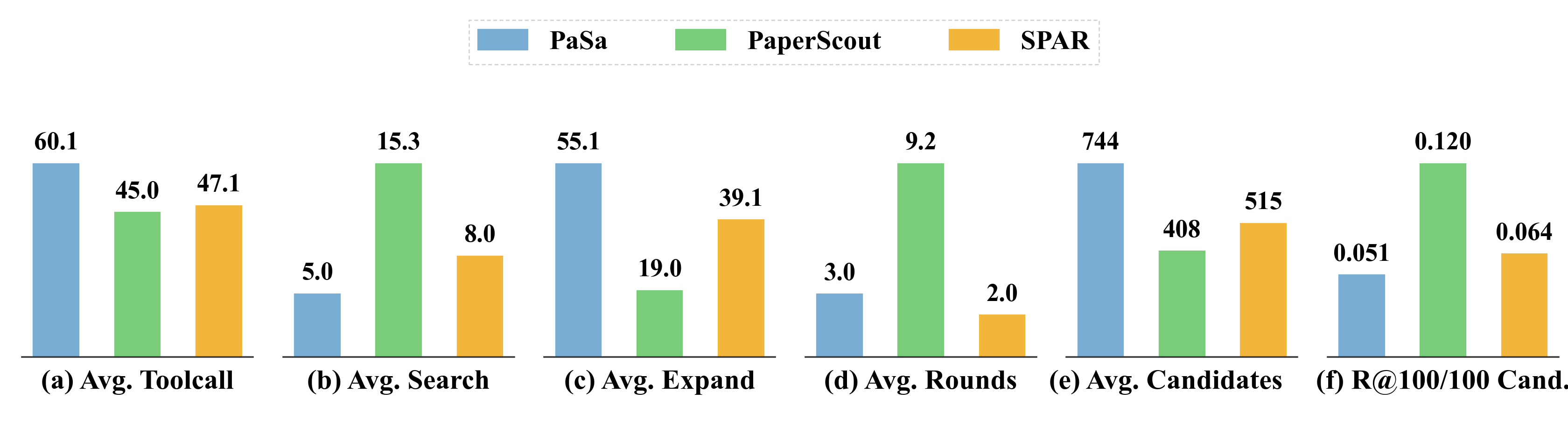}
    \caption{Efficiency and process statistics of paper search agents, including number of tool calls, rounds, observed candidates, and R@100/100 Cand., which normalizes Recall@100 by every 100 observed candidates.}
    \label{fig:search_efficiency}
\end{figure}

\subsection{Main Results}

We evaluate search performance at two levels: system-level evaluation compares complete literature search systems on ScholarQuest, while model-level evaluation compares retrieval quality and Planner token usage on ScholarQuest-100 under a shared search harness.

\subsubsection{System-level Evaluation}
\label{sec:system_level_results}

Table~\ref{tab:main_table} reports Recall@25, Recall@100, and Recall@All across four research intents and the overall query set of ScholarQuest.
Traditional retrieval methods show clear limitations on complex paper search queries, especially on compositional constraints, as shown by the low R@100 of Google Search (0.006) and Google Scholar (0.010) on \textit{scope-controlled} queries.
\rev{Dense and Hybrid retrieval improve over sparse retrieval, achieving 0.206 and 0.212 overall R@100.} However, they remain single-shot retrievers and cannot iteratively refine queries, expand promising papers, or integrate intermediate evidence. \rev{In contrast, agentic methods with multi-round decision-making achieve higher overall R@100 scores, ranging from 0.269 to 0.312.}

%
%


The results also show clear differences among agentic search methods. PaperScout achieves 0.408 R@100 on \textit{method-oriented}, outperforming PaSa and SPAR by 0.063 and 0.075, respectively.
It also leads on \textit{setting-anchored} and \textit{comparison-based}, with R@100 scores of 0.341 and 0.327, higher than PaSa (0.296, 0.294) and SPAR (0.282, 0.280). One possible explanation is that more adaptive multi-turn search can help agents better navigate scattered literature evidence. We conduct further analysis in the next section to examine tool-use patterns, search efficiency, and exploration behavior.

\subsubsection{Model-level Evaluation}
\label{sec:model_level_results}

Figure~\ref{fig:model_level_results} compares seven model configurations on ScholarQuest-100, with limits of 15 interaction rounds and five parallel tool calls per round. Overall, performance varies across the four research intents: most models show relatively similar performance on \textit{setting-anchored} queries, whereas differences are more pronounced on \textit{method-oriented} and \textit{comparison-based} queries. Performance on \textit{scope-controlled} queries is generally low, indicating that fine-grained control over retrieval scope remains challenging.

Specifically, GPT-5.6-Sol achieves the highest Recall@All across all four intents, reaching 0.648 overall. However, its recall is 0.817 on setting-anchored queries but only 0.405 on scope-controlled queries, highlighting the persistent difficulty of scope control even for the strongest-performing configuration. Other models exhibit distinct strengths: Kimi-K2.5 outperforms Gemini 3.1 Pro on setting-anchored and method-oriented queries, while Gemini performs better on comparison-based and scope-controlled queries. The task-trained PaperScout-4B also remains competitive despite its compact size, achieving 0.241 recall on scope-controlled queries and outperforming Qwen3.8-Max, Kimi-K2.5, and GLM-5.3. Moreover, greater token consumption does not consistently translate into better retrieval coverage: GPT-5.6-Sol uses 62.8k mean Planner tokens per query, compared with Gemini's 68.6k, while achieving higher recall.

\begin{figure}[!t]
    \centering
    \begin{subfigure}[b]{0.49\linewidth}
        \centering
        \includegraphics[width=\linewidth]{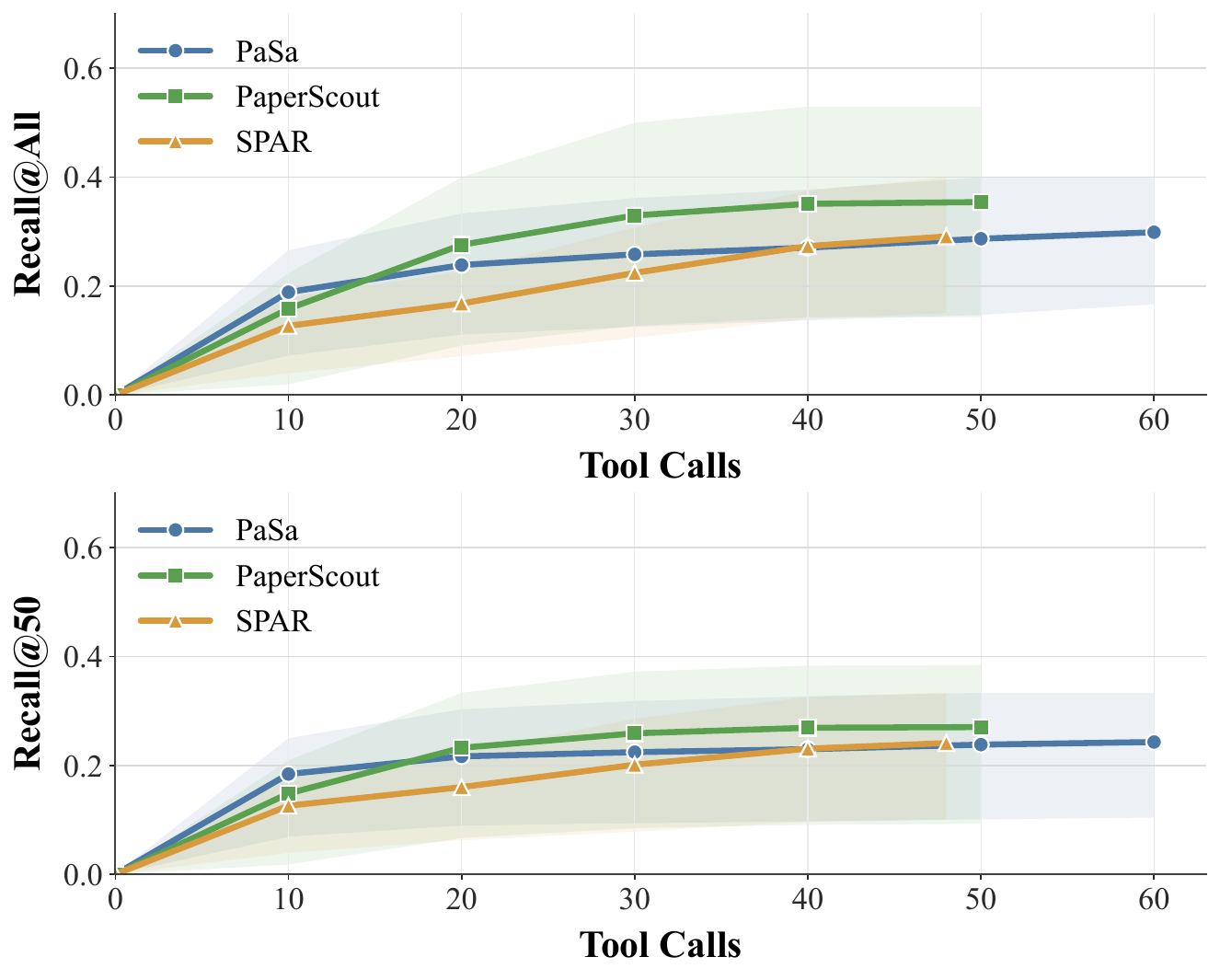}
        \caption{System-level}
        \label{fig:toolcall_system}
    \end{subfigure}\hfill
    \begin{subfigure}[b]{0.49\linewidth}
        \centering
        \includegraphics[width=\linewidth]{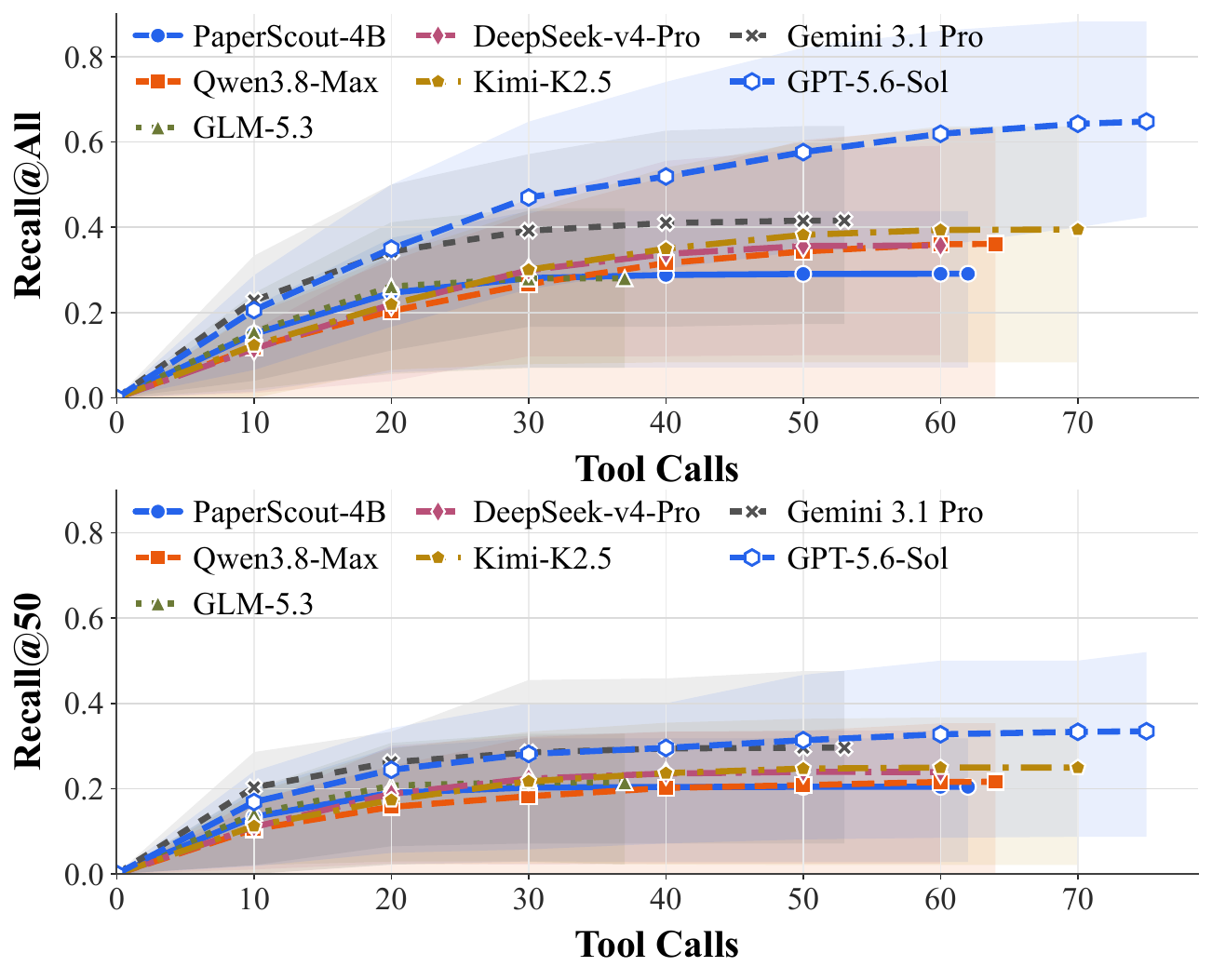}
        \caption{Model-level}
        \label{fig:toolcall_model}
    \end{subfigure}
    \captionsetup{font=small}
    \caption{Recall trajectories for (a) system-level evaluation on ScholarQuest and (b) model-level evaluation on ScholarQuest-100. The upper and lower panels show \mbox{Recall@All} and \mbox{Recall@50}, respectively. Shaded regions indicate the 25th--75th percentile range across queries.}
    \label{fig:toolcall}
\end{figure}

\Needspace{5\baselineskip}
\subsection{Analysis of Agentic Search Behavior}
\label{sec:agentic_behavior}

\rev{We examine search efficiency, coverage growth, and robustness across query conditions, connecting complete search systems with models under a shared harness.}


\paragraph{Search Efficiency and Coverage Growth.}
\rev{PaperScout uses fewer tool calls and examines fewer candidates than PaSa and SPAR (Figure~\ref{fig:search_efficiency}), yet achieves the highest overall Recall@All (Table~\ref{tab:main_table}). Its Recall@100 per 100 observed candidates is 0.120, versus 0.051 for PaSa and 0.064 for SPAR. In the reconstructed trajectories (Figure~\ref{fig:toolcall}(\subref{fig:toolcall_system})), PaperScout reaches higher coverage in fewer steps than PaSa and a higher final recall than SPAR. We hypothesize that PaperScout's process rewards, which encourage relevance gains from new papers and penalize repeated tool calls~\citep{pan2026paperscout}, improve the yield of intermediate retrieval actions. This provides a direct learning signal at each interaction, linking tool choices to the relevant evidence they uncover. The resulting policy may use accumulated results to decide when to pursue a new search direction and which retrieved papers to expand. Combined with optimization of cumulative future returns, this feedback may guide adaptive use of search and citation expansion, improving overall planning and coverage efficiency.}

\rev{Under the shared harness, GPT-5.6-Sol averages 64.3 tool calls and approximately 1,095 candidate papers per valid query, achieving a Recall@All of 0.648. Gemini 3.1 Pro averages 32.4 calls and approximately 419 candidates, with a Recall@All of 0.415. GPT also shows greater efficiency in candidate discovery, yielding 17.0 new candidates per Search call compared with Gemini's 7.0. Specifically, GPT favors short, anchor-focused queries and explores multiple keyword formulations; its queries average 3.0 words, compared with Gemini's 5.0. This combination of concise anchors and varied keyword formulations may support more productive candidate discovery, helping explain GPT's higher Recall@All (Figure~\ref{fig:toolcall}(\subref{fig:toolcall_model})).}

\paragraph{Robustness Across Query Conditions.}
\rev{Scope-controlled queries challenge all three search systems in handling exclusion constraints (Appendix~\ref{app:agentic_search_analysis}). Similar difficulties arise in model comparisons under the shared harness. Even GPT-5.6-Sol, which leads in overall performance, achieves a Recall@All of only 0.405 on scope-controlled queries, substantially below its 0.817 on setting-anchored queries. This gap indicates persistent challenges in handling complex scope constraints.}

\rev{One possible explanation is that setting-anchored queries provide explicit positive cues, such as datasets or experimental settings, that help models locate candidate papers and assess relevance. Scope-controlled queries instead require distinguishing relevant papers from semantically similar work outside the specified boundaries. Models may misinterpret exclusion terms as positive search cues, directing retrieval toward papers that should be excluded. Subsequent citation expansion may follow these related topics and further divert search from the intended scope. These queries therefore place greater demands on preserving constraints across search rounds and checking candidates against them, which may help explain their lower recall.}

\subsection{Failure Analysis}
\label{sec:failure_analysis}

\vspace{-0.05in}

\begin{figure}[t]
    \centering
    \includegraphics[width=\linewidth]{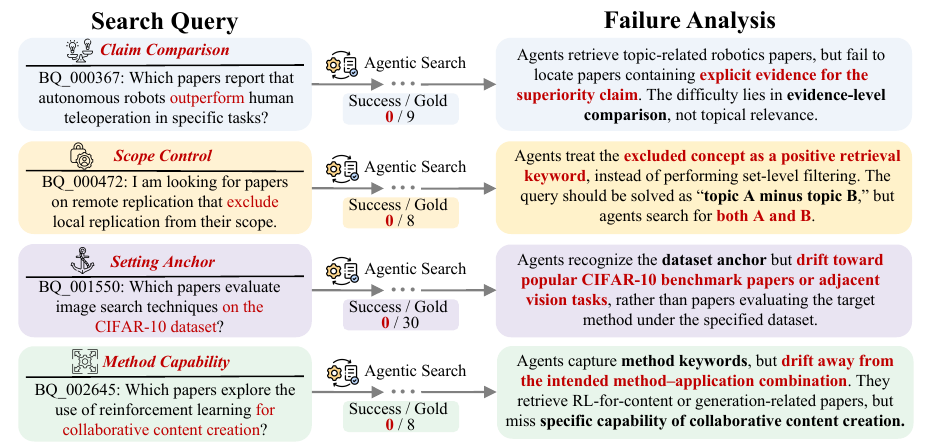}
    \caption{Representative common zero-recall cases across query types. Each case shows a query where all three agents retrieve no gold papers, together with the corresponding gold-answer size and the main failure cause.}
    \label{fig:failure_case}
\end{figure}

\rev{Figure~\ref{fig:failure_case} presents representative examples for each query type, showing that failures arise from different forms of intent mismatch rather than a uniform retrieval error.
\textbf{For comparison-based queries}, agents retrieve related papers but miss evidence for the specific comparison.
In the autonomous-robot case, they fail to find papers explicitly showing autonomous robots \textit{outperform} human teleoperation, yielding 0 out of 9 gold papers.
\textbf{For scope-controlled queries}, agents struggle with negative constraints: in the remote replication case, the target is essentially ``remote replication minus local replication,'' but agents treat the excluded concept as a positive keyword, yielding 0 out of 8 gold papers.
\textbf{For setting-anchored queries}, agents recognize anchors such as CIFAR-10 or Pascal VOC, but drift toward popular benchmark papers instead of papers evaluating the target method under the specified setting.
\textbf{For method-oriented queries}, agents match the method keyword but fail to preserve compositional intent, such as reinforcement learning for collaborative content creation.
These failures highlight a gap between topical relevance and intent satisfaction: semantically related papers may still fail to satisfy the required comparison, exclusion constraint, or method--setting combination.
Further retrieval or citation expansion within a plausible but off-target neighborhood may reinforce that direction rather than recover the missing evidence.
Together with the results in Table~\ref{tab:zero_recall_cases} (Appendix~\ref{app:failure_statistics}), these cases highlight a core requirement: agents should not only search broadly, but also target the right evidence, constraints, and method--setting combinations.}

\vspace{-0.05in}

\section{Conclusion}


\vspace{-0.1in}

We propose ScholarQuest, a taxonomy-guided benchmark with broad CS topic coverage, four research intents, scalable answer construction, and a shared \textit{ScholarBase} environment for reproducible agentic paper search evaluation.
\rev{Agentic methods outperform single-shot retrieval: the best system improves overall R@100 from 0.212 to 0.312 over the strongest non-agentic baseline, a 47.2\% relative gain. System- and model-level analyses distinguish broad coverage from shortlist quality and zero-recall robustness, showing that greater search effort does not consistently yield better evidence. ScholarQuest provides a diagnostic testbed for intent preservation, constraint-aware exploration, and evidence-level retrieval, supporting the development of reliable and efficient academic search agents.}

\section{Limitations}

ScholarQuest is designed as a controlled benchmark for agentic paper search, but it still has several boundaries. First, ScholarQuest focuses on computer science topics and uses an arXiv-grounded literature environment, so it does not cover the full diversity of scholarly communication across disciplines, venues, and publication formats. Second, relevance judgments are based on paper titles, abstracts, and metadata rather than full-text evidence, which makes the benchmark scalable but may miss fine-grained claims that only appear in the body of a paper. Finally, although our answer construction pipeline combines multi-source retrieval, citation expansion, and LLM-based relevance adjudication, automatic construction may still miss some relevant papers in open literature environments. These limitations motivate future extensions toward broader literature sources, full-text-aware relevance assessment, and stronger answer-set validation.

\section*{Reproducibility Statement}

To support reproducibility, we provide the ScholarQuest dataset, the \textit{\textbf{ScholarBase}} retrieval backend, and the shared evaluation harness. The main text and appendix describe the data construction pipeline, experimental configurations, and metric definitions, with implementation details in Appendix~\ref{app:implementation_details} and prompt templates in Appendix~\ref{app:prompts}. These resources are available through the public repository linked in the abstract. Fixed evaluation sets and a common retrieval environment support reproducible comparisons across methods.

\section*{Ethics Statement}

This study evaluates academic paper retrieval using titles, abstracts, metadata, and citation relations from publicly available scholarly literature. Researchers participating in manual review and dataset curation are compensated, as detailed in Appendix~\ref{app:annotator_compensation}. We use human review and independent adjudication to reduce automated labeling errors, while acknowledging potential biases in literature coverage and query generation. Relevance labels indicate how well a paper matches a specific query and are not intended to assess the quality of individual papers or the abilities of researchers.

\section*{AI Use Statement}

We use generative AI for writing assistance, literature retrieval and discovery, and synthetic dataset construction. For writing, AI assists with drafting selected passages, translation, and language polishing. For retrieval and discovery, AI supports query rewriting and candidate paper discovery, as described in Section~\ref{sec:answer_discovery}. For dataset construction, Qwen3-Max maps ACM CCS topics to arXiv categories and generates queries across four research intents, while LLM-based relevance judges produce initial answer labels (Sections~\ref{sec:query_generation} and~\ref{sec:answer_discovery}). Prompt templates are provided in Appendix~\ref{app:prompts}. Data quality is controlled through the filtering, human auditing, and iterative curation procedures described in the paper. The authors take responsibility for reviewing and verifying AI-assisted content, as well as for the final manuscript and research outputs.

\bibliography{scholarquest}
\bibliographystyle{iclr2027_conference}

\appendix

\clearpage

\section{Implementation Details}
\label{app:implementation_details}

We provide reproducibility details for ScholarQuest here. We organize the details into benchmark construction and dataset evaluation.

\subsection{Details of Benchmark Construction}
\label{app:construction_dedup}

\paragraph{Query Construction.}
\rev{We start from 1,682 ACM CCS topics and use Qwen3-Max to map each topic to one or more arXiv subject categories. Only topics assigned to arXiv CS categories in Table~\ref{tab:arxiv_cs_categories} are retained, resulting in a construction pool of 1,631 CS taxonomy nodes. We draw 20,000 topic-seed instances with replacement using arXiv category-frequency weights, deterministically shuffle these draws, and select 10,000 instances for query generation. Of these, 9,972 complete successfully; the remaining 28 consist of eight batch errors and 20 parse errors. For each successful sampled seed instance, Qwen3-Max generates exactly four queries, one for each intent type: method-oriented, setting-anchored, comparison-based, and scope-controlled, yielding 39,888 raw query records. We structurally validate each record, deduplicate identical \texttt{final\_query} strings, and retain queries with 5--200 final strict-match answers, yielding 1,111 benchmark queries.}

\begin{wraptable}{r}{0.52\textwidth}
\centering
\small
\vspace{-0.15in}
\caption{Default configuration for ScholarQuest benchmark construction.}
\setlength{\tabcolsep}{4pt}
\resizebox{\linewidth}{!}{
\begin{tabular}{lc}
\toprule
\textbf{Configuration} & \textbf{Value} \\
\midrule
Raw ACM CCS topics & 1,682 \\
\rev{Retained CS taxonomy nodes} & \rev{1,631} \\
Final queries & 1,111 \\
\rev{Queries per successful instance} & \rev{4} \\
Query rewrites per query & 10 \\
Search top-$k$ per rewrite & 10 \\
Retrieval sources & Google Search, arXiv, Semantic Scholar \\
Query-level concurrency & 4 \\
Citation expansion depth & Up to 2 hops \\
Citation limit per seed paper & 30 \\
Reference limit per seed paper & All available references \\
First-stage filtering & Recall-oriented prefilter \\
Relevance scoring rubric & 0--2 \\
Answer relevance threshold & 2.0 \\
Maximum answer count & 500 \\
Release answer-set filter & 5--200 strict matches \\
Maximum scored papers per query & 10,000 \\
Relevance judges & Multiple LLM-based judges \\
\bottomrule
\end{tabular}
}
\label{tab:construction_config}
\end{wraptable}

\paragraph{Answer Construction.}
For each query, we generate 10 rewritten search queries and retrieve the top 10 results for each rewrite from complementary retrieval sources, including Google Search, arXiv, and Semantic Scholar. These services use different indexes and ranking mechanisms and therefore act as complementary discovery channels; arXiv is the target literature space used for normalization and evaluation. Every matched arXiv ID is canonicalized by removing its version suffix (e.g., \texttt{v2}). Records found through different sources, rewrites, or later citation-expansion paths are merged by the normalized base arXiv ID, preserving a single paper identity throughout construction. We use a recall-oriented prefilter to remove clear mismatches, followed by LLM-based relevance adjudication using the query, paper title, abstract and metadata. Papers judged as strict matches are treated as primary answer candidates.

For citation expansion, each high-confidence seed paper retrieves up to 30 citing papers and all available references. We further perform second-hop expansion from high-confidence first-hop neighbors while pruning noisy branches. Newly discovered papers are normalized, deduplicated, and scored with the same relevance-filtering pipeline. The expansion stops when no eligible high-confidence seed remains, the answer set reaches 500 papers, or the number of scored unique papers reaches 10,000 for a query.

\begin{wraptable}{r}{0.48\textwidth}
\centering
\small
\vspace{-0.15in}
\caption{Confusion matrix between LLM-based relevance scores and human annotations. Hit Rate denotes the exact agreement rate within each LLM score group.}
\setlength{\tabcolsep}{7pt}
\renewcommand{\arraystretch}{1.08}
\resizebox{\linewidth}{!}{
\begin{tabular}{lccccc}
\toprule
\multirow{2}{*}{\textbf{LLM Score}}
& \multicolumn{3}{c}{\textbf{Human Score}}
& \multirow{2}{*}{\textbf{Total}}
& \multirow{2}{*}{\textbf{Hit Rate}} \\
\cmidrule(lr){2-4}
& \textbf{0} & \textbf{1} & \textbf{2} & & \\
\midrule
0 & 133 & 16  & 1   & 150 & 88.7\% \\
1 & 21  & 117 & 12  & 150 & 78.0\% \\
2 & 2   & 19  & 129 & 150 & 86.0\% \\
\midrule
\textbf{Total} & 156 & 152 & 142 & 450 & 84.2\% \\
\bottomrule
\end{tabular}
}
\label{tab:llm_human_confusion}
\end{wraptable}

\paragraph{Human Audit.}
To assess label quality without requiring exhaustive manual annotation, we conduct a targeted human audit of the final-stage LLM relevance scores. We sample 150 query-paper pairs from each score group, resulting in 450 pairs in total, and ask three Ph.D. experts to re-score them based on the user query, paper title, abstract, and metadata. Human annotators follow the same three-level rubric as the LLM judges: 0 = mismatch, 1 = partial match, and 2 = strict match.
For each pair, we aggregate the three expert annotations by majority vote and use this aggregated label as the human score in the following agreement analysis.
As shown in Table~\ref{tab:llm_human_confusion}, LLM scores align well with human annotations, with Pearson correlation of 0.867 \citep{pearson1896vii}, Spearman correlation of 0.867 \citep{spearman1961proof}, and quadratic weighted Cohen's $\kappa$ of 0.866 \citep{cohen1968weighted}. The score-2 group used for final answer selection reaches 86.0\% strict-match precision (129/150) and 98.7\% relaxed precision (148/150 with human score $\geq 1$). In contrast, only 1/150 score-0 candidates and 12/150 score-1 candidates are judged as strict matches by humans, suggesting that residual false-negative risk is concentrated in borderline cases.

Overall, these results indicate that the automated construction pipeline can reliably separate strict-match answers from mismatches, with residual errors mainly concentrated in borderline cases. These findings motivate the additional adjudication of borderline and high-disagreement cases. We further present four representative disagreement cases in Table~\ref{tab:human_llm_disagreement}, illustrating typical failure modes of LLM-based relevance scoring, such as over-reliance on topical overlap and insufficient sensitivity to fine-grained query constraints.


\subsection{Iterative Curation and Convergence of ScholarQuest-100}
\label{app:curation_convergence}

ScholarQuest-100 contains 100 queries, with 25 queries per research intent, and receives the additional iterative review described in Section~\ref{sec:human_loop_curation}. Its quality-control objective has two parts: removing papers that fail the query's constraints and discovering relevant papers missing from the current answer set. We distinguish these query-level checks from the 450-pair audit above, which assesses the reliability of automatically assigned relevance labels in the full ScholarQuest benchmark.

\paragraph{Query Boundaries and Evidence Standard.}
For each query, the review specifies the necessary topic, method, setting, comparison direction, and exclusion conditions, as applicable. Query refinement clarifies ambiguous wording while preserving the underlying research intent. A substantive change to an inclusion or exclusion condition defines a revised query version and requires reassessment of affected candidates. Answer coverage is assessed relative to each fixed query version. Throughout a review cycle, paper eligibility is defined relative to the same frozen literature environment, and relevance follows the title--abstract--metadata evidence standard used by the benchmark. A paper qualifies as a strict match only when the available evidence supports all necessary constraints.

\paragraph{Discovery and Reassessment Rounds.}
Each complete round combines a relevance review with a search for omissions. The relevance review checks retained answers and borderline candidates against the current query definition, removes unsupported inclusions, and resolves conflicting judgments. The discovery stage uses complementary query formulations, terminology variants, and citation/reference expansion from verified relevant papers to seek additional candidates in the frozen environment. Newly discovered candidates undergo the same relevance assessment as existing answers. A complete round finishes the planned discovery actions and assesses all resulting candidates. If a query constraint changes, the affected judgments are revisited before convergence is assessed for the revised version.

\paragraph{Operational Convergence.}
Convergence is assessed separately for each query and its answer set. A convergence decision requires a complete discovery-and-reassessment round satisfying all of the following conditions:
\begin{enumerate}[leftmargin=*, itemsep=2pt, topsep=3pt]
    \item \textbf{Stable query interpretation.} The necessary constraints and relevance boundary remain unchanged, with no unresolved ambiguity that could alter inclusion decisions.
    \item \textbf{Stable relevance judgments.} Review yields no further changes to the retained answer set or the grades of previously judged candidates. Disagreements affecting answer inclusion have been adjudicated, and unresolved cases are not promoted to strict matches.
    \item \textbf{No newly verified strict matches.} The completed supplementary searches and citation checks produce no additional strict-match papers after normalization and deduplication. All candidates returned by these actions have been assessed, and no planned discovery action or newly identified expansion lead remains pending within the review scope.
\end{enumerate}

\paragraph{Interpreting Precision and Answer Coverage.}
The two evaluation sets provide complementary forms of quality assurance. For the 1,111-query ScholarQuest benchmark, the stratified human audit provides empirical evidence about automatic label reliability: the score-2 sample has 86.0\% strict-match precision and 98.7\% relaxed precision (Table~\ref{tab:llm_human_confusion}). The strict value corresponds to the benchmark's answer-inclusion standard, while the relaxed value captures topical or partial relevance. This audit evaluates label quality among the scored candidates.

For ScholarQuest-100, constraint-level reassessment targets false inclusions and therefore supports precision, while complementary discovery targets omissions and therefore supports recall. Operational convergence combines stable relevance decisions with empirical saturation of answer discovery under a fixed query, evidence standard, and literature environment.

\begingroup
\color{black}
\subsection{Topic Coverage and Query Selection}
\label{app:query_validity}

\paragraph{Final Topic Coverage.}
We distinguish the size of the taxonomy used to construct ScholarQuest from the topic coverage of the released benchmark. The construction pipeline begins with 1,682 ACM CCS leaf nodes and retains 1,631 nodes mapped to at least one arXiv CS category. To approximate the empirical prevalence of arXiv subfields, it then draws 20,000 topic-seed instances with replacement using category-frequency weights. These draws cover 1,487 unique ACM leaf nodes. The draws are deterministically shuffled, and 10,000 instances covering 1,317 nodes are selected for query generation. Of these, 9,972 complete successfully while retaining coverage of all 1,317 nodes. Exact-string query deduplication leaves 13,097 query records associated with 1,308 nodes. The final release contains 1,111 queries associated with 528 sampled seed instances, 373 unique ACM leaf nodes, 355 distinct topic labels, and 36 reported arXiv CS categories. Thus, ``1,631 topics'' describes the available construction pool, whereas the final benchmark itself spans 373 taxonomy nodes. We use the latter quantity when describing the topic coverage of the released evaluation set.

\begin{center}
\begin{minipage}{\linewidth}
\centering
\small
\captionsetup{labelfont={color=black},textfont={color=black}}
\captionof{table}{Topic and query coverage across ScholarQuest construction stages. Rows 1--5 count taxonomy nodes or sampled seed instances; rows 6--8 count query records. Sampling with replacement allows seed-instance counts to exceed unique ACM leaf nodes.}
\label{tab:query_construction_funnel}
\setlength{\tabcolsep}{5pt}
\renewcommand{\arraystretch}{1.08}
\resizebox{0.74\linewidth}{!}{
\begin{tabular}{lrr}
\toprule
\textbf{Construction stage} & \textbf{Records / draws} & \textbf{Unique ACM nodes} \\
\midrule
Raw ACM CCS leaf nodes & 1,682 & 1,682 \\
Retained CS taxonomy nodes & 1,631 & 1,631 \\
Weighted topic-seed draws & 20,000 & 1,487 \\
Selected generation instances & 10,000 & 1,317 \\
Successful generation instances & 9,972 & 1,317 \\
Generated query records & 39,888 & 1,317 \\
After exact-string deduplication & 13,097 & 1,308 \\
Final benchmark release & 1,111 & 373 \\
\bottomrule
\end{tabular}
}
\end{minipage}
\end{center}

\paragraph{Query Filtering Funnel.}
Query construction proceeds through topic-seed sampling, query generation, exact-string deduplication, and answer construction and filtering. Of the 10,000 selected generation instances, 9,972 complete successfully, yielding 39,888 raw query records. Exact-string deduplication removes 26,791 records (67.2\%), leaving 13,097 unique queries. The subsequent reduction to 1,111 released queries reflects query selection for answer construction, execution outcomes, and the requirement of 5--200 strict-match answers. Table~\ref{tab:query_construction_funnel} summarizes the recorded stage-level counts.

\paragraph{Selection Effects.}
We compare the intent distributions of the deduplicated query pool and the released benchmark. In the 13,097-query pool, method-oriented, setting-anchored, scope-controlled, and comparison-based queries account for 23.3\%, 23.9\%, 21.9\%, and 30.9\%, respectively. Their corresponding shares in the final set are 27.2\% (302), 29.0\% (322), 28.5\% (317), and 15.3\% (170). Thus, the share of comparison-based queries decreases from 30.9\% to 15.3\% across the construction pipeline. The 5--200-answer criterion defines the answer-set range of the released evaluation task, and these distributions characterize its composition across research intents.

\endgroup

\subsection{Details of Dataset Evaluation}
\label{app:evaluation_protocol}

We compare nine baselines from three groups: standard retrieval methods, external academic search systems, and agentic search methods.

\begin{itemize} 
    \item \textbf{Dense Retrieval.} Dense retriever using BGE-M3 title--abstract embeddings.
    \item \textbf{Hybrid Retrieval.} Hybrid retriever combining BM25 and dense retrieval with RRF.
    \item \textbf{Google Search.} A general web search baseline for paper discovery.
    \item \textbf{Google Scholar.} A widely used academic search engine baseline.
    \item \textbf{Semantic Scholar.} An academic search API with paper metadata and citation information.
    \item \textbf{DeepXiv.} An external scientific literature search interface.
    \item \textbf{PaSa.} LLM-based agent for multi-step academic paper search.
    \item \textbf{SPAR.} Training-free workflow with retrieval, citation exploration, and reranking.
    \item \textbf{PaperScout.} Autonomous paper search agent that adaptively decides when and how to search or expand.
\end{itemize}


\paragraph{Retrieval Baselines.}
We evaluate two \textit{ScholarBase} retrieval baselines: dense retrieval and RRF-based hybrid retrieval. Dense retrieval ranks papers with BGE-M3 title--abstract embeddings. Hybrid retrieval combines sparse and dense rankings through reciprocal rank fusion. Both methods take the original query as input and return ranked papers from \textit{ScholarBase}.

\paragraph{Academic Search Systems.}
We evaluate four external academic search systems: Google Search, Google Scholar, and search APIs of Semantic Scholar and DeepXiv. For each query, each system returns up to 300 papers. Returned papers are matched to ScholarQuest answers through arXiv identifiers, titles, and available metadata, ensuring that all methods are evaluated against the same arXiv-grounded answer sets.

\paragraph{Agentic Search Methods.}
We evaluate three agentic paper search systems: PaSa, SPAR, and PaperScout. To ensure a fair comparison, we run their released inference code and open checkpoints without modifying their model weights or decision logic. All tool interfaces are provided by \textit{\textbf{ScholarBase}}, including paper search, metadata lookup, and citation/reference expansion, so that different agents interact with the same literature backend. During evaluation, all retrieved or expanded papers are normalized by arXiv ID before metric computation. We also record process statistics, including interaction rounds, search calls, expansion calls, observed candidates, and Recall@100 per 100 observed candidates.

\paragraph{Ranking Normalization and Metrics.}
Before evaluation, returned papers are mapped to normalized base arXiv IDs, version suffixes are removed, and duplicate papers are collapsed while preserving their earliest rank. Let $L_q=(p_1,\ldots,p_{m_q})$ be this normalized unique ranking for query $q$, let $G_q$ be its gold set, and let $L_q^{(k)}$ be the first $\min(k,m_q)$ papers. Query-level recall is
\begin{equation}
R_q@k = \frac{|L_q^{(k)}\cap G_q|}{|G_q|}.
\end{equation}
At the same normalized output cutoff, query-level Precision and F1 are
\begin{equation}
\begin{aligned}
P_q@k &= \frac{|L_q^{(k)}\cap G_q|}{|L_q^{(k)}|}, \\
F1_q@k &= \frac{2P_q@kR_q@k}{P_q@k+R_q@k}.
\end{aligned}
\end{equation}
where each quantity is set to zero when its denominator is zero. Recall, Precision, and F1 are first computed per query and then macro-averaged. We report Precision@25, Precision@100, F1@25, and F1@100 as complements to the recall-oriented main results.
Recall@All evaluates the complete normalized unique ranking returned at system termination. Each system operates under its original search, expansion, interaction-round, and stopping budgets.

\begin{table*}[t]
\centering
\small
\caption{Query-level macro Precision (top) and F1 (bottom) at common normalized output cutoffs, grouped by query type. Bold indicates the best result within each query type and metric.}
\label{tab:app_precision_f1}
\renewcommand{\arraystretch}{1.02}

\begin{tabular*}{0.98\textwidth}{@{\extracolsep{\fill}}lcccccccc@{}}
\toprule
\multirow{2}{*}{\textbf{Method}}
& \multicolumn{2}{c}{\textbf{Method-oriented}}
& \multicolumn{2}{c}{\textbf{Setting-anchored}}
& \multicolumn{2}{c}{\textbf{Scope-controlled}}
& \multicolumn{2}{c}{\textbf{Comparison-based}} \\
\cmidrule(lr){2-3}\cmidrule(lr){4-5}\cmidrule(lr){6-7}\cmidrule(lr){8-9}
& \textbf{P@25} & \textbf{P@100}
& \textbf{P@25} & \textbf{P@100}
& \textbf{P@25} & \textbf{P@100}
& \textbf{P@25} & \textbf{P@100} \\
\midrule
Dense & 0.211 & 0.121 & 0.141 & 0.080 & 0.059 & 0.045 & 0.063 & 0.029 \\
Hybrid & 0.221 & 0.124 & 0.150 & 0.082 & 0.080 & 0.048 & 0.064 & 0.030 \\
Google Search & 0.181 & 0.079 & 0.086 & 0.046 & 0.012 & 0.006 & 0.037 & 0.012 \\
Google Scholar & 0.120 & 0.049 & 0.100 & 0.042 & 0.008 & 0.006 & 0.054 & 0.018 \\
Semantic Scholar & 0.091 & 0.037 & 0.141 & 0.048 & 0.002 & 0.001 & 0.045 & 0.015 \\
DeepXiv & 0.130 & 0.094 & 0.093 & 0.075 & 0.014 & 0.014 & 0.020 & 0.017 \\
\midrule
PaSa & 0.358 & 0.159 & 0.323 & 0.128 & \textbf{0.216} & \textbf{0.120} & 0.136 & 0.045 \\
SPAR & 0.356 & 0.153 & 0.310 & 0.118 & 0.209 & 0.116 & 0.126 & 0.041 \\
PaperScout & \textbf{0.369} & \textbf{0.178} & \textbf{0.330} & \textbf{0.150} & 0.181 & 0.114 & \textbf{0.144} & \textbf{0.051} \\
\bottomrule
\end{tabular*}

\vspace{6pt}
\begin{tabular*}{0.98\textwidth}{@{\extracolsep{\fill}}lcccccccc@{}}
\toprule
\multirow{2}{*}{\textbf{Method}}
& \multicolumn{2}{c}{\textbf{Method-oriented}}
& \multicolumn{2}{c}{\textbf{Setting-anchored}}
& \multicolumn{2}{c}{\textbf{Scope-controlled}}
& \multicolumn{2}{c}{\textbf{Comparison-based}} \\
\cmidrule(lr){2-3}\cmidrule(lr){4-5}\cmidrule(lr){6-7}\cmidrule(lr){8-9}
& \textbf{F1@25} & \textbf{F1@100}
& \textbf{F1@25} & \textbf{F1@100}
& \textbf{F1@25} & \textbf{F1@100}
& \textbf{F1@25} & \textbf{F1@100} \\
\midrule
Dense & 0.144 & 0.145 & 0.102 & 0.099 & 0.032 & 0.049 & 0.076 & 0.048 \\
Hybrid & 0.147 & 0.148 & 0.107 & 0.102 & 0.042 & 0.052 & 0.075 & 0.050 \\
Google Search & 0.112 & 0.087 & 0.058 & 0.054 & 0.005 & 0.006 & 0.055 & 0.022 \\
Google Scholar & 0.081 & 0.057 & 0.070 & 0.051 & 0.004 & 0.006 & 0.062 & 0.028 \\
Semantic Scholar & 0.061 & 0.044 & 0.098 & 0.058 & 0.001 & 0.001 & 0.051 & 0.023 \\
DeepXiv & 0.087 & 0.110 & 0.065 & 0.090 & 0.008 & 0.015 & 0.023 & 0.026 \\
\midrule
PaSa & 0.237 & 0.184 & 0.221 & 0.150 & \textbf{0.114} & \textbf{0.124} & 0.152 & 0.069 \\
SPAR & 0.235 & 0.177 & 0.213 & 0.140 & 0.112 & 0.121 & 0.144 & 0.063 \\
PaperScout & \textbf{0.258} & \textbf{0.208} & \textbf{0.234} & \textbf{0.176} & 0.097 & 0.118 & \textbf{0.161} & \textbf{0.077} \\
\bottomrule
\end{tabular*}
\end{table*}

\section{Complementary Evaluation Results}
\label{app:complementary_evaluation}

\paragraph{Complementary Precision and F1 Results.}
Table~\ref{tab:app_precision_f1} reports the complete complementary results. Across the four query types, the agentic methods outperform the non-agentic retrieval and external-search baselines on all reported Precision/F1 metrics. PaperScout leads 12 of the 16 query-type-specific metric cells, attaining the best Precision@25, Precision@100, F1@25, and F1@100 on method-oriented, setting-anchored, and comparison-based queries. PaSa leads all four complementary metrics on scope-controlled queries. The stronger F1 scores indicate that agentic methods not only retrieve more relevant papers but also maintain higher precision than non-agentic methods.

\paragraph{Repeated Runs and Confidence Intervals.}
The non-agentic methods are largely deterministic under the fixed index, dataset, and retrieval configuration. PaSa, SPAR, and PaperScout are each evaluated in three independent runs over all 1,111 queries, and Table~\ref{tab:main_table} reports their mean performance. Table~\ref{tab:app_agent_recall_stability} reports the corresponding run-level standard deviations, which are small across query types and Recall metrics.

\begin{figure}[t]
\centering
\includegraphics[width=\textwidth]{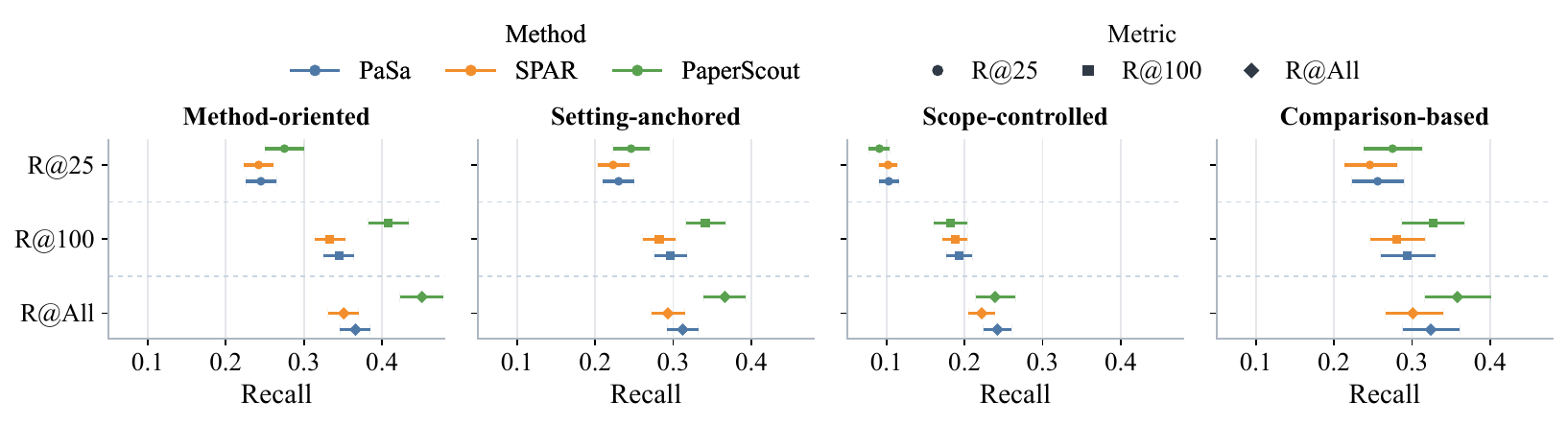}
\caption{Agent Recall point estimates and 95\% query-bootstrap confidence intervals across query types. Points denote Recall estimates and horizontal whiskers denote confidence intervals. Dashed horizontal lines separate Recall@25, Recall@100, and Recall@All.}
\label{fig:app_agent_recall_ci}
\end{figure}

\begin{table*}[t]
\centering
\small
\caption{Average Recall@100 of agentic paper search methods across gold answer-set size buckets and query types. PaperScout is strongest on method and setting queries, while scope-control remains difficult and recall decreases as answer sets grow.}
\setlength{\tabcolsep}{4pt}
\renewcommand{\arraystretch}{1.05}
\resizebox{\textwidth}{!}{
\begin{tabular}{lccccccccccccccc}
\toprule
\textbf{Answer Size}
& \multicolumn{3}{c}{\textbf{Method-oriented}}
& \multicolumn{3}{c}{\textbf{Setting-anchored}}
& \multicolumn{3}{c}{\textbf{Scope-controlled}}
& \multicolumn{3}{c}{\textbf{Comparison-based}}
& \multicolumn{3}{c}{\textbf{Overall}} \\
\cmidrule(lr){1-1}
\cmidrule(lr){2-4}
\cmidrule(lr){5-7}
\cmidrule(lr){8-10}
\cmidrule(lr){11-13}
\cmidrule(lr){14-16}
& \textbf{PaSa} & \textbf{SPAR} & \textbf{PaperScout}
& \textbf{PaSa} & \textbf{SPAR} & \textbf{PaperScout}
& \textbf{PaSa} & \textbf{SPAR} & \textbf{PaperScout}
& \textbf{PaSa} & \textbf{SPAR} & \textbf{PaperScout}
& \textbf{PaSa} & \textbf{SPAR} & \textbf{PaperScout} \\
\midrule
5--25
& 0.401 & 0.390 & \textbf{0.512}
& 0.342 & 0.337 & \textbf{0.389}
& \textbf{0.268} & 0.252 & 0.252
& 0.311 & 0.300 & \textbf{0.342}
& 0.338 & 0.327 & \textbf{0.387} \\

26--50
& 0.346 & 0.340 & \textbf{0.413}
& 0.312 & 0.296 & \textbf{0.370}
& 0.214 & \textbf{0.219} & 0.190
& 0.235 & 0.230 & \textbf{0.294}
& 0.295 & 0.288 & \textbf{0.337} \\

51--100
& 0.320 & 0.299 & \textbf{0.344}
& 0.241 & 0.216 & \textbf{0.285}
& 0.147 & 0.149 & \textbf{0.152}
& 0.182 & 0.139 & \textbf{0.213}
& 0.229 & 0.215 & \textbf{0.251} \\

101--150
& 0.257 & 0.249 & \textbf{0.262}
& 0.242 & 0.213 & \textbf{0.262}
& \textbf{0.168} & 0.166 & 0.143
& \textbf{0.183} & 0.127 & 0.173
& \textbf{0.214} & 0.202 & 0.209 \\

151--200
& 0.238 & 0.237 & \textbf{0.243}
& 0.175 & 0.159 & \textbf{0.195}
& \textbf{0.144} & 0.130 & 0.142
& 0.165 & 0.110 & \textbf{0.177}
& 0.176 & 0.164 & \textbf{0.181} \\
\bottomrule
\end{tabular}
}
\label{tab:app_recall100_by_size_type}
\end{table*}

\begin{table*}[t]
\centering
\scriptsize
\caption{Run-level standard deviations of agent Recall over three independent runs. The corresponding mean Recall values are reported in Table~\ref{tab:main_table}.}
\label{tab:app_agent_recall_stability}
\setlength{\tabcolsep}{3pt}
\renewcommand{\arraystretch}{1.05}
\resizebox{\textwidth}{!}{
\begin{tabular}{lcccccccccccc}
\toprule
\multirow{2}{*}{\textbf{Method}}
& \multicolumn{3}{c}{\textbf{Method-oriented}}
& \multicolumn{3}{c}{\textbf{Setting-anchored}}
& \multicolumn{3}{c}{\textbf{Scope-controlled}}
& \multicolumn{3}{c}{\textbf{Comparison-based}} \\
\cmidrule(lr){2-4}\cmidrule(lr){5-7}\cmidrule(lr){8-10}\cmidrule(lr){11-13}
& \textbf{R@25} & \textbf{R@100} & \textbf{R@All}
& \textbf{R@25} & \textbf{R@100} & \textbf{R@All}
& \textbf{R@25} & \textbf{R@100} & \textbf{R@All}
& \textbf{R@25} & \textbf{R@100} & \textbf{R@All} \\
\midrule
PaSa & 0.0047 & 0.0031 & 0.0032 & 0.0034 & 0.0034 & 0.0020 & 0.0020 & 0.0030 & 0.0019 & 0.0038 & 0.0047 & 0.0028 \\
SPAR & 0.0033 & 0.0035 & 0.0010 & 0.0022 & 0.0028 & 0.0007 & 0.0016 & 0.0024 & 0.0007 & 0.0031 & 0.0038 & 0.0010 \\
PaperScout & 0.0024 & 0.0029 & 0.0007 & 0.0020 & 0.0026 & 0.0011 & 0.0017 & 0.0031 & 0.0006 & 0.0029 & 0.0036 & 0.0010 \\
\bottomrule
\end{tabular}
}
\end{table*}

To quantify uncertainty from benchmark composition, we also compute 95\% query-bootstrap confidence intervals within each intent category. For a category with $n$ queries, we sample $n$ queries with replacement, recompute macro Recall, and repeat this procedure 10,000 times; the 2.5th and 97.5th percentiles define the interval. Figure~\ref{fig:app_agent_recall_ci} shows that the bootstrap intervals preserve the main trends while separating query-sampling uncertainty from run-to-run variability.

\subsection{\rev{Additional Analyses of Agentic Search}}
\label{app:agentic_search_analysis}

\paragraph{Robustness Across Research Intents.}
\rev{Agentic methods generally outperform traditional retrieval methods across metrics.
However, their performance varies substantially across research intents, revealing important limitations of current paper search agents.
As shown in Table~\ref{tab:main_table}, all three agents perform relatively well on \textit{method-oriented}, \textit{setting-anchored}, and \textit{comparison-based} queries, where the search intent is mainly guided by positive semantic cues such as methods, tasks, settings, or comparative claims.
In contrast, \textit{scope-controlled} is consistently the most challenging category: PaSa, SPAR, and PaperScout achieve only 0.193, 0.188, and 0.182 R@100, respectively, which are clearly lower than their performance on other intents.}

\begin{wrapfigure}{R}{0.57\textwidth}
    \vspace{-\intextsep}
    \centering
    \includegraphics[width=\linewidth]{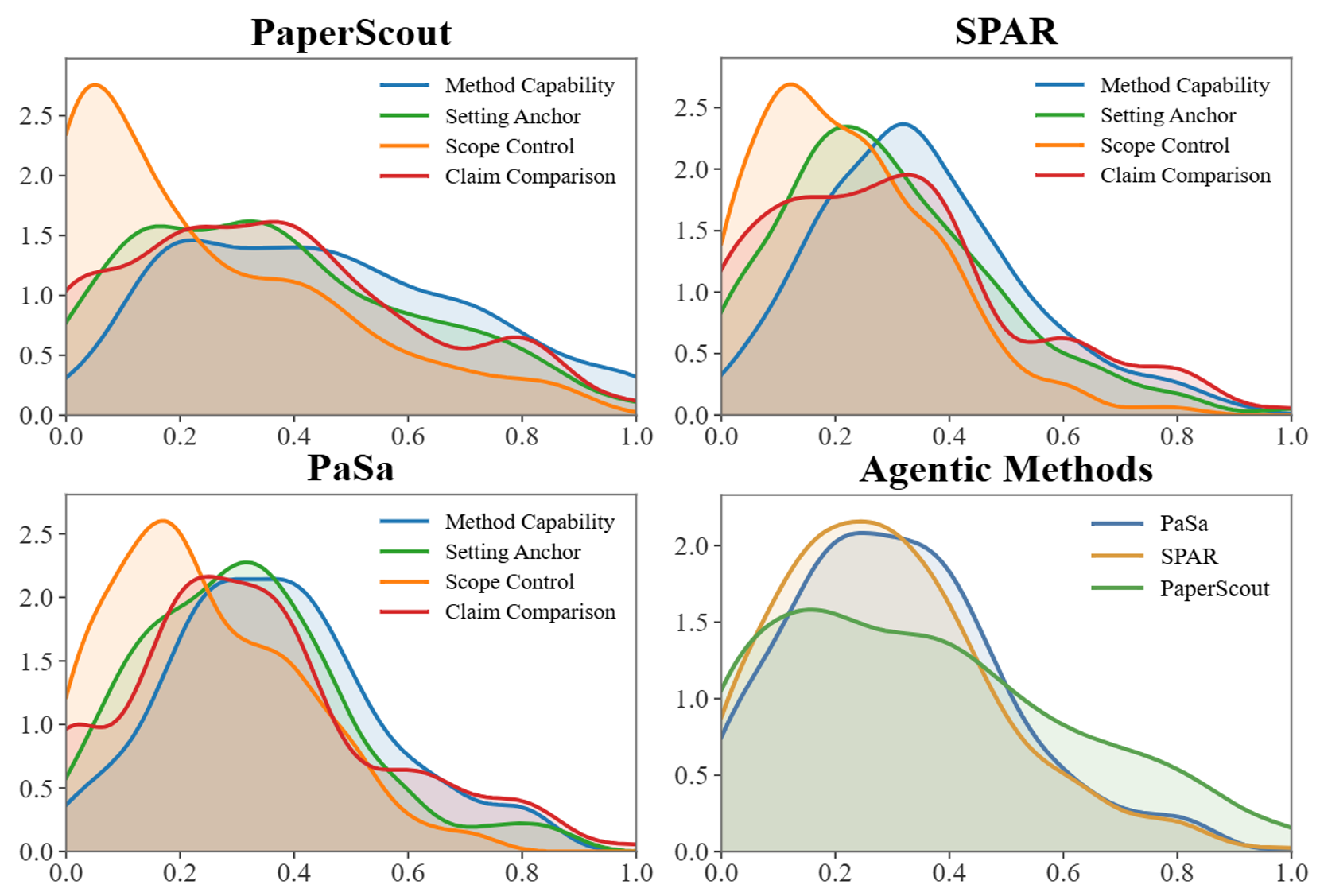}
    \captionsetup{font=small}
    \caption{\rev{Recall@All density distributions across agentic literature search methods. Panels show research-intent distributions for each method and the overall comparison.}}
    \label{fig:recall}
    \label{fig:retrieval_analysis}
\end{wrapfigure}

\rev{Figure~\ref{fig:recall} further provides a distributional view of this limitation.
For all three agents, the recall distributions of \textit{scope-controlled} queries are concentrated in the low-recall region, indicating that this category is not only worse on average but also more prone to query-level failures.
Unlike positive-intent queries, scope-control queries require agents to retrieve papers that match the main topic while preserving exclusion constraints or fine-grained scope boundaries during multi-round exploration.
Although PaperScout shows a more right-shifted overall recall distribution, its scope-control curve remains heavily skewed toward low recall.
These results suggest that adaptive tool use improves robustness, but constraint-sensitive paper search remains a shared limitation of agentic methods.}

\begin{wraptable}{r}{0.56\textwidth}
\vspace{-\intextsep}
\centering
\small
\captionsetup{font=small}
\caption{Average Recall@All across gold answer-set size buckets. \#Query denotes the number of queries.}
\label{tab:answer_size}
\setlength{\tabcolsep}{4pt}
\renewcommand{\arraystretch}{1.05}

\resizebox{\linewidth}{!}{%
\begin{tabular}{lcccc}
\toprule
\multirow{2}{*}{\textbf{Answer Size}}
& \multirow{2}{*}{\textbf{\#Query}}
& \multicolumn{3}{c}{\textbf{Recall@All}} \\
\cmidrule(lr){3-5}
&
& \textbf{PaSa}
& \textbf{SPAR}
& \textbf{PaperScout} \\
\midrule
5--25    & 457 & 0.352 & 0.338 & \textbf{0.405} \\
26--50   & 218 & 0.319 & 0.307 & \textbf{0.371} \\
51--100  & 225 & 0.267 & 0.242 & \textbf{0.304} \\
101--150 & 130 & 0.265 & 0.239 & \textbf{0.281} \\
151--200 & 81  & 0.235 & 0.205 & \textbf{0.278} \\
\bottomrule
\end{tabular}%
}

\end{wraptable}

\paragraph{Robustness Across Answer-set Sizes.}
\rev{We evaluate robustness by grouping queries by the number of ground-truth answers.
As shown in Table~\ref{tab:answer_size}, average recall decreases for all agents as answer sets grow, suggesting that broad-answer queries remain challenging under limited search budgets.
For instance, PaSa drops from 0.352 in the 5--25 bucket to 0.235 in the 151--200 bucket, while SPAR drops from 0.338 to 0.205.
PaperScout achieves the highest average recall in every bucket, with scores of 0.405, 0.371, 0.304, 0.281, and 0.278 from small to large answer sets.
Its advantage persists even in the largest bucket, where it outperforms PaSa and SPAR by 0.043 and 0.073 average recall, respectively.
These results indicate that PaperScout better balances focused exploration and broad coverage across varying answer-set sizes.
Detailed results are shown in Table~\ref{tab:app_recall100_by_size_type}.}

\subsection{Model-level Comparison Protocol}
\label{app:model_level_protocol}

Figure~\ref{fig:model_level_results} uses model-specific valid query sets. Queries with failed or incomplete results or invalid token records are excluded from both the recall and token means; failures are not assigned zero recall. Valid counts are 100/100/99/93/100/94/94 for PaperScout-4B, Qwen3.8-Max, GLM-5.3, DeepSeek-v4-Pro, Kimi-K2.5, Gemini 3.1 Pro, and GPT-5.6-Sol, respectively. GPT and Gemini each have six failed queries, leaving 25/25/23/21 valid queries across setting-anchored, method-oriented, comparison-based, and scope-controlled intents, although their failed query IDs differ. Overall means weight intent-level means by their valid query counts.

Token usage includes only reported Planner input and output from the final successful attempt, with reasoning counted once as part of completion tokens. Failed attempts, preflight requests, and Selector usage are excluded. For Gemini, we sum the reported prompt and completion components; discrepancies of at most one token in the redundant API total are audited without excluding otherwise complete traces. Other configurations retain strict token-total validation. These are descriptive comparisons across configurations with different valid query sets, tokenizers, and inference settings.

\Needspace{7\baselineskip}
\subsection{Model-level Behavioral Diagnostics}
\label{app:matched_behavior}

\rev{Table~\ref{tab:matched_behavior} summarizes the completed runs of all seven model configurations on ScholarQuest-100, using the same results as Figures~\ref{fig:model_level_results} and~\ref{fig:toolcall}(\subref{fig:toolcall_model}) under the protocol in Appendix~\ref{app:model_level_protocol}. For each query, we compute Recall@All and Recall@50 from the final ranking and gold answers, count actual tool calls from the interaction logs, and reconstruct the unique candidate pool from final paper provenance.}

\begin{table}[!ht]
\centering
\small
\caption{Behavioral diagnostics for completed model runs on ScholarQuest-100. Calls and reconstructed unique candidate pools are query means; Expand is the fraction of all logged calls assigned to citation expansion.}
\label{tab:matched_behavior}
\setlength{\tabcolsep}{4pt}
\begin{tabular*}{\linewidth}{@{\extracolsep{\fill}}lrrrrr@{}}
\toprule
Model & Calls & Expand (\%) & Candidates & R@All & R@50 \\
\midrule
PaperScout-4B & 45.8 & 44.2 & 418 & 0.291 & 0.204 \\
Qwen3.8-Max & 38.6 & 76.0 & 629 & 0.361 & 0.216 \\
GLM-5.3 & 51.2 & 24.2 & 319 & 0.280 & 0.215 \\
DeepSeek-v4-Pro & 60.6 & 29.5 & 463 & 0.357 & 0.239 \\
Kimi-K2.5 & 49.3 & 64.2 & 645 & 0.395 & 0.249 \\
Gemini 3.1 Pro & 32.4 & 45.4 & 419 & 0.415 & 0.296 \\
GPT-5.6-Sol & 64.3 & 49.0 & 1,095 & 0.648 & 0.334 \\
\bottomrule
\end{tabular*}
\end{table}

\paragraph{Tool Allocation.}
\rev{The configurations differ substantially in how they divide their tool budget between search and citation expansion. Qwen3.8-Max and Kimi-K2.5 devote 76.0\% and 64.2\% of calls to expansion, respectively, and construct relatively large candidate pools of 629 and 645 papers. GLM-5.3 and DeepSeek-v4-Pro favor search, with expansion shares of 24.2\% and 29.5\% and smaller pools of 319 and 463 papers. However, the largest pool belongs to GPT-5.6-Sol, whose 49.0\% expansion share is close to an even allocation. These patterns suggest that expansion frequency alone does not determine coverage: the choice of search queries and expansion seeds, together with the total search effort, also matters. The statistics count all logged calls, including calls yielding no new papers; the trajectory steps in Figure~\ref{fig:toolcall} instead represent nonempty source groups reconstructed from final provenance.}

\paragraph{Search Effort and Candidate Accumulation.}
\rev{More calls do not consistently produce a larger or more useful candidate pool. Gemini 3.1 Pro makes the fewest calls (32.4), 29.1\% fewer than PaperScout-4B, while reaching nearly the same pool size (419 versus 418). Nevertheless, its Recall@All is higher (0.415 versus 0.291), as is its Recall@50 (0.296 versus 0.204), suggesting differences in which papers are discovered and retained near the top of the ranking. Qwen also constructs a larger pool than DeepSeek with fewer calls (629 versus 463 candidates from 38.6 versus 60.6 calls). Dividing mean pool size by mean call count gives approximately 17.0 unique candidates per call for GPT, 16.3 for Qwen, and 6.2 for GLM. These ratios summarize net candidate accumulation across the run, including empty and duplicate-result calls; they do not measure the isolated yield of either tool. Thus, call count and pool size provide complementary views of search effort, while recall measures whether that effort recovers relevant evidence.}

\paragraph{Coverage and Shortlist Quality.}
\rev{The benefit of broader exploration is substantially smaller at the shortlist cutoff. GPT makes approximately twice as many calls as Gemini and builds a candidate pool 2.6 times as large. Its Recall@All advantage is 0.2324, but its Recall@50 advantage is only 0.0385. Applying the decomposition $R@\mathrm{All}=R@50+(R@\mathrm{All}-R@50)$ to the reported macro means places 83.4\% of this coverage advantage beyond the first 50 results. GPT's mean recall outside the top 50 is 0.313, compared with 0.120 for Gemini. The ordering of configurations can also change with the cutoff: Qwen exceeds DeepSeek in Recall@All (0.361 versus 0.357), but trails it in Recall@50 (0.216 versus 0.239). Kimi improves on Qwen at both cutoffs, yet remains below Gemini despite its larger pool. These observations distinguish discovering additional relevant papers from placing them in a short list that users can inspect. Because all configurations use the shared ranker, the metrics reflect the interaction between each discovered pool and that ranker; they do not isolate the Planner's ranking ability.}

\Needspace{5\baselineskip}
\paragraph{Tool Calls and Planner Tokens.}
\rev{Tool-call counts and Planner token usage capture different resource demands. GPT makes slightly more calls than DeepSeek (64.3 versus 60.6), but consumes fewer mean Planner tokens (62.8k versus 81.4k, a 22.8\% reduction) while achieving higher recall at both cutoffs. Conversely, Gemini uses the fewest calls, yet its mean Planner usage of 68.6k tokens exceeds Qwen's 51.3k and Kimi's 55.5k. A tool call therefore cannot be treated as a fixed unit of language-model computation: input context length, generated output, and reasoning usage contribute separately to token consumption. These figures describe the observed configurations under their respective tokenizers and inference settings, including GPT's dedicated keyword-search prompt, rather than isolating an effect of model identity.}

\paragraph{Intent-Dependent Token Tradeoffs.}
\rev{Figure~\ref{fig:model_level_results} further shows that aggregate comparisons hide differences across research intents. On method-oriented queries, Qwen improves mean Recall@All from 0.332 to 0.462 over PaperScout-4B (39.1\% relative gain), while mean Planner tokens increase only from 60.6k to 62.6k (3.3\%). Kimi extends coverage to 0.535 with 66.3k tokens, a further 15.7\% recall gain over Qwen for 5.9\% more tokens. Their ordering changes on comparison-based queries, where Qwen reaches 0.332 and Kimi 0.310; Gemini reaches 0.471, accompanied by higher token usage (69.3k versus Kimi's 46.4k). GPT achieves the highest reported recall in every intent, but its scope-controlled recall remains much lower than its setting-anchored recall (0.405 versus 0.817). Together, these results motivate evaluating tool allocation, token consumption, and shortlist quality by research intent, rather than assuming that one global resource budget or exploration strategy is equally effective across query types.}

\Needspace{0.30\textheight}
\section{\rev{Zero-recall Failure Statistics}}
\label{app:failure_statistics}
\label{app:failure_analysis}

\begin{wraptable}{r}{0.60\textwidth}
\vspace{-\intextsep}
\centering
\small
\captionsetup{font=small}
\caption{Common \textbf{zero-recall} cases by query type. \#Failures denotes the number of queries where all three agents retrieve no gold papers; Avg. Candidates reports the average number of accessed candidate papers.}
\label{tab:zero_recall_cases}
\setlength{\tabcolsep}{4pt}
\renewcommand{\arraystretch}{1.05}

\resizebox{\linewidth}{!}{%
\begin{tabular}{lcccc}
\toprule
\multirow{2}{*}{\textbf{Query Type}}
& \multirow{2}{*}{\textbf{\#Failures}}
& \multicolumn{3}{c}{\textbf{Avg. Candidates}} \\
\cmidrule(lr){3-5}
&
& \textbf{PaSa}
& \textbf{SPAR}
& \textbf{PaperScout} \\
\midrule
Comparison-based & 7 & 926.6  & 589.9 & 405.7 \\
Scope-controlled & 6 & 871.2  & 618.0 & 315.3 \\
Setting-anchored & 5 & 713.6  & 551.8 & 472.8 \\
Method-oriented  & 2 & 1306.0 & 831.0 & 532.0 \\
\midrule
Overall          & 20 & 894.6 & 612.9 & 408.0 \\
\bottomrule
\end{tabular}%
}

\end{wraptable}

\rev{We analyze common zero-recall cases, where PaSa, SPAR, and PaperScout all fail to retrieve any gold paper. As shown in Table~\ref{tab:zero_recall_cases}, such failures appear across all four query types. Although these cases are rare, covering only 20 out of 1,111 complete agentic queries (1.80\%), they are highly diagnostic: all three agents access a substantial number of candidate papers on these queries, yet none of the retrieved papers match the gold answers.}

\paragraph{Interpretation.}
\rev{These cases highlight the role of search direction, query intent, and constraint preservation in locating relevant evidence through retrieval and citation expansion.}

\rev{Representative cases are analyzed in Section~\ref{sec:failure_analysis}.}

\section{Tool-call Evidence for Agentic Search}
\label{app:tool_call_evidence}

We further inspect the tool-use trajectory of \texttt{BQ\_002897}, the method-capability case shown in Appendix~\ref{app:case_studies}. The query asks for papers that explore deep reinforcement learning for video summarization. This case is useful because the gold answers form a compact but nontrivial neighborhood: some answers are directly reachable through targeted keyword search, while others are only recovered through citation/reference expansion. \rev{Under the benchmark's title-and-abstract evidence standard, the final answer set for this case contains 11 papers.}

Table~\ref{tab:bq002897_tool_chain} lists every tool call that directly increases cumulative recall. \rev{PaperScout reaches all 11 gold answers after 20 tool calls.} \rev{Its first four search calls recover 9/11 answers, and two later expansion calls recover the remaining two answers.} \rev{In contrast, SPAR and PaSa stop at 8/11 and 7/11, respectively.} \rev{The missing answers are also systematic: both SPAR and PaSa miss \texttt{2109.01309}, \texttt{1807.09418}, and \texttt{2007.14552}, while PaSa additionally misses \texttt{2105.06441}.}

Table~\ref{tab:bq002897_gold_routes} gives a gold-answer-level view of the same query. It shows that PaperScout's advantage does not come from more direct hits alone: it uses early successful searches as anchors and then follows expansion paths from relevant seeds. \rev{For example, expanding \texttt{1801.00054} recovers \texttt{2105.06441}, while expanding \texttt{1805.02792} recovers \texttt{1807.09418}.} This explains why only a small number of tool calls change recall: marginal gold discoveries are sparse, and many additional calls inspect already relevant neighborhoods or provide post-saturation confirmation rather than adding new gold answers.

\section{Annotator Compensation}
\label{app:annotator_compensation}
All human annotators are compensated for their work. The compensation rate is set to no less than ten times the local minimum hourly wage for each participant, reflecting both the specialized expertise required and the time-intensive nature of benchmark refinement and evaluation.

\begin{table}[t]
\centering
\scriptsize
\caption{Evidence chain for \texttt{BQ\_002897}. Each row is a tool call that directly increases cumulative recall. PaperScout first locates most gold answers through targeted search, then uses expansion to recover the remaining answers.}
\setlength{\tabcolsep}{2pt}
\renewcommand{\arraystretch}{1.05}
\begin{tabular}{p{0.08\linewidth}p{0.06\linewidth}p{0.08\linewidth}p{0.38\linewidth}p{0.25\linewidth}p{0.08\linewidth}}
\toprule
\textbf{Method} & \textbf{Call} & \textbf{Tool} & \textbf{Query or Expansion Seed} & \textbf{Newly Found Gold Answers} & \textbf{Recall} \\
\midrule
PaperScout & 1 & search & deep reinforcement learning AND video summarization & \texttt{2001.05864} & \rev{1/11} \\
PaperScout & 2 & search & reinforcement learning AND video summarization & \texttt{2007.14552}, \texttt{2106.10528}, \texttt{2109.01309} & \rev{4/11} \\
PaperScout & 3 & search & deep reinforcement learning AND action selection AND video summarization & \texttt{1801.00054}, \texttt{1807.03089}, \texttt{2005.09531}, \texttt{2407.04258} & \rev{8/11} \\
PaperScout & 4 & search & reinforcement learning AND keyframe selection AND video summarization & \texttt{2505.23268} & \rev{9/11} \\
PaperScout & 6 & expand & \texttt{1801.00054} & \rev{\texttt{2105.06441}} & \rev{10/11} \\
PaperScout & 20 & expand & \texttt{1805.02792} & \texttt{1807.09418} & \rev{11/11} \\
\midrule
SPAR & 1 & search & Literature review of application domains in deep reinforcement learning for video summarization & \texttt{2407.04258} & \rev{1/11} \\
SPAR & 2 & expand & \texttt{2405.08890} & \texttt{1801.00054}, \texttt{2001.05864} & \rev{3/11} \\
SPAR & 3 & expand & \texttt{2101.06072} & \texttt{1807.03089}, \texttt{2005.09531} & \rev{5/11} \\
SPAR & 7 & expand & \texttt{2407.04258} & \texttt{2505.23268} & \rev{6/11} \\
SPAR & 14 & search & State-of-the-art in deep reinforcement learning for video summarization & \texttt{2105.06441} & \rev{7/11} \\
SPAR & 32 & expand & \texttt{2412.08357} & \texttt{2106.10528} & \rev{8/11} \\
\midrule
PaSa & 1 & search & Survey papers on video summarization using deep reinforcement learning & \texttt{1801.00054}, \texttt{1807.03089}, \texttt{2005.09531}, \texttt{2407.04258}, \texttt{2505.23268} & \rev{5/11} \\
PaSa & 8 & expand & \texttt{2101.06072} & \texttt{2001.05864} & \rev{6/11} \\
PaSa & 10 & expand & \texttt{2410.04449} & \texttt{2106.10528} & \rev{7/11} \\
\bottomrule
\end{tabular}

\label{tab:bq002897_tool_chain}
\end{table}

\begin{table}[t]
\centering
\small
\caption{Full names of arXiv computer science subject categories. These categories are used to map ACM CCS topics into the CS topic space for ScholarQuest construction.}
\setlength{\tabcolsep}{6pt}
\resizebox{0.98\textwidth}{!}{
\begin{tabular}{ll|ll}
\toprule
\textbf{Category} & \textbf{Full Name}
& \textbf{Category} & \textbf{Full Name} \\
\midrule
cs.AI & Artificial Intelligence
& cs.AR & Hardware Architecture \\
cs.CC & Computational Complexity
& cs.CE & Computational Engineering, Finance, and Science \\
cs.CG & Computational Geometry
& cs.CL & Computation and Language \\
cs.CR & Cryptography and Security
& cs.CV & Computer Vision and Pattern Recognition \\
cs.CY & Computers and Society
& cs.DB & Databases \\
cs.DC & Distributed, Parallel, and Cluster Computing
& cs.DL & Digital Libraries \\
cs.DM & Discrete Mathematics
& cs.DS & Data Structures and Algorithms \\
cs.ET & Emerging Technologies
& cs.FL & Formal Languages and Automata Theory \\
cs.GL & General Literature
& cs.GR & Graphics \\
cs.GT & Computer Science and Game Theory
& cs.HC & Human-Computer Interaction \\
cs.IR & Information Retrieval
& cs.IT & Information Theory \\
cs.LG & Machine Learning
& cs.LO & Logic in Computer Science \\
cs.MA & Multiagent Systems
& cs.MM & Multimedia \\
cs.MS & Mathematical Software
& cs.NA & Numerical Analysis \\
cs.NE & Neural and Evolutionary Computing
& cs.NI & Networking and Internet Architecture \\
cs.OH & Other Computer Science
& cs.OS & Operating Systems \\
cs.PF & Performance
& cs.PL & Programming Languages \\
cs.RO & Robotics
& cs.SC & Symbolic Computation \\
cs.SD & Sound
& cs.SE & Software Engineering \\
cs.SI & Social and Information Networks
& cs.SY & Systems and Control \\
\bottomrule
\end{tabular}
}

\label{tab:arxiv_cs_categories}
\end{table}

\begin{table}[t]
\centering
\scriptsize
\setlength{\tabcolsep}{4pt}
\renewcommand{\arraystretch}{1.12}
\caption{
Representative disagreement cases between human judgments and LLM-generated labels.
Human labels are obtained through manual review, while LLM labels correspond to the original automatic relevance labels.
Scores 0, 1, and 2 denote mismatch, partial match, and strict match, respectively.
}
\label{tab:human_llm_disagreement}
\resizebox{\textwidth}{!}{
\begin{tabular}{p{0.26\textwidth} p{0.40\textwidth} p{0.29\textwidth}}
\toprule
\textbf{Query and Label} & \textbf{Paper Title and Abstract Evidence} & \textbf{LLM Label Bias} \\
\midrule

\textbf{Query:} Which papers evaluate network domains in the context of data center networks? \newline
\textbf{Human:} 0 (mismatch) \newline
\textbf{LLM:} 1 (partial match)
&
\textbf{Title:} Efficient Coflow Scheduling in Hybrid-Switched Data Center Networks \newline
\textbf{Evidence:} The paper studies coflow scheduling in hybrid-switched data center networks, focusing on scheduling, communication time, hybrid links, and performance guarantees. However, it does not explicitly evaluate network domains.
&
\textbf{Keyword over-crediting.}
The LLM is attracted by the strong topic match ``data center networks,'' but overlooks the missing constraint of ``network domains,'' leading to an overestimated relevance label. \\

\midrule

\textbf{Query:} Which papers report that machine learning-based information extraction outperforms rule-based systems? \newline
\textbf{Human:} 1 (partial match) \newline
\textbf{LLM:} 0 (mismatch)
&
\textbf{Title:} PAM: Understanding Product Images in Cross Product Category Attribute Extraction \newline
\textbf{Evidence:} The paper studies product attribute extraction with a transformer-based model using product text, OCR tokens, and visual objects. It is related to machine learning-based information extraction, but does not directly report outperforming rule-based systems.
&
\textbf{Over-strict claim matching.}
The LLM treats the comparison with rule-based systems as mandatory, while human reviewers still regard the paper as a partial match due to its clear task-level match. \\

\midrule

\textbf{Query:} Which papers study image search using deep learning techniques? \newline
\textbf{Human:} 1 (partial match) \newline
\textbf{LLM:} 2 (strict match)
&
\textbf{Title:} Visual Discovery at Pinterest \newline
\textbf{Evidence:} The paper presents Pinterest's visual discovery engine for visual search and recommendation products. It mentions object detection and improved engagement, but does not explicitly state ``deep learning'' or specify neural architectures.
&
\textbf{Implicit-knowledge overuse.}
The LLM likely infers deep learning from visual search and object detection, upgrading the paper to a strict match without explicit textual evidence for the method constraint. \\

\midrule

\textbf{Query:} Which papers claim that modern imaging techniques outperform traditional methods? \newline
\textbf{Human:} 2 (strict match) \newline
\textbf{LLM:} 1 (partial match)
&
\textbf{Title:} Twin-beam sub-shot-noise raster-scanning microscope \newline
\textbf{Evidence:} The paper proposes a quantum imaging microscope and reports improved precision over a shot-noise-limited classical version while preserving resolution and optical power.
&
\textbf{Terminology mismatch.}
The LLM underestimates relevance because the paper expresses ``traditional methods'' as a domain-specific ``classical version,'' whereas human reviewers recognize the comparison as sufficient evidence. \\

\bottomrule
\end{tabular}
}
\end{table}


\begin{table}[t]
\centering
\scriptsize
\caption{Gold-answer provenance for \texttt{BQ\_002897}. S\# and E\# denote the order of search and expansion calls, respectively, through which a method discovers each answer.}
\setlength{\tabcolsep}{2pt}
\renewcommand{\arraystretch}{1.05}
\begin{tabular}{p{0.10\linewidth}p{0.29\linewidth}p{0.22\linewidth}p{0.16\linewidth}p{0.16\linewidth}}
\toprule
\textbf{Gold Answer} & \textbf{Title} & \textbf{PaperScout Route} & \textbf{SPAR Route} & \textbf{PaSa Route} \\
\midrule
\texttt{2505.23268} & Unsupervised Transcript-assisted Video Summarization and Highlight Detection & S4 search: keyframe selection query & E7 expand \texttt{2407.04258} & S1 search: survey query \\
\texttt{2407.04258} & Reinforcement Learning for Unsupervised Video Summarization With Reward Generator Training & S3 search: action-selection query & S1 search: literature-review query & S1 search: survey query \\
\texttt{2109.01309} & Unsupervised multi-latent space reinforcement learning framework for video summarization in ultrasound imaging & S2 search: reinforcement learning query & Missed & Missed \\
\texttt{2106.10528} & Video Summarization Through Reinforcement Learning With a 3D Spatio-Temporal U-Net & S2 search: reinforcement learning query & E32 expand \texttt{2412.08357} & E10 expand \texttt{2410.04449} \\
\texttt{2105.06441} & DeepQAMVS: Query-Aware Hierarchical Pointer Networks for Multi-Video Summarization & E6 expand \texttt{1801.00054} & S14 search: state-of-the-art query & Missed \\
\texttt{2007.14552} & Compare and Select: Video Summarization with Multi-Agent Reinforcement Learning & S2 search: reinforcement learning query & Missed & Missed \\
\texttt{2005.09531} & Ultrasound Video Summarization using Deep Reinforcement Learning & S3 search: action-selection query & E3 expand \texttt{2101.06072} & S1 search: survey query \\
\texttt{2001.05864} & Weakly Supervised Video Summarization by Hierarchical Reinforcement Learning & S1 search: deep reinforcement learning query & E2 expand \texttt{2405.08890} & E8 expand \texttt{2101.06072} \\
\texttt{1807.09418} & Video Storytelling: Textual Summaries for Events & E20 expand \texttt{1805.02792} & Missed & Missed \\
\texttt{1807.03089} & Video Summarisation by Classification with Deep Reinforcement Learning & S3 search: action-selection query & E3 expand \texttt{2101.06072} & S1 search: survey query \\
\texttt{1801.00054} & Deep Reinforcement Learning for Unsupervised Video Summarization with Diversity-Representativeness Reward & S3 search: action-selection query & E2 expand \texttt{2405.08890} & S1 search: survey query \\
\bottomrule
\end{tabular}

\label{tab:bq002897_gold_routes}
\end{table}

\clearpage
\onecolumn

\section{Case Studies}
\label{app:case_studies}

We provide four representative ScholarQuest cases, one for each query intent. Each case shows the query metadata and its complete gold-answer set sorted in descending order by the first arXiv submission date.

\begin{tcolorbox}[
breakable,
title={Case 1: Method-Capability Query (\texttt{BQ\_002897})},
colback=gray!5,
colframe=gray!50!black,
fonttitle=\bfseries\small,
before skip=4pt,
after skip=6pt
]
\small
\textbf{Query.} Which papers explore the use of deep reinforcement learning for video summarization?\\
\rev{\textbf{Domain.} \texttt{cs.CV}; \textbf{Topic seed.} Video summarization; \textbf{Constraint.} \texttt{technique = deep reinforcement learning}; \textbf{Gold answers.} 11.}

\vspace{3pt}
\scriptsize
\setlength{\tabcolsep}{1pt}
\renewcommand{\arraystretch}{1.02}
\begin{tabular}{@{}p{0.16\linewidth}@{\hspace{1.5pt}}p{0.17\linewidth}@{\hspace{2pt}}p{0.63\linewidth}@{}}
\toprule
\textbf{Date} & \textbf{arXiv ID} & \textbf{Title} \\
\midrule
2025-05-29 & \texttt{2505.23268} & Unsupervised Transcript-assisted Video Summarization and Highlight Detection \\
2024-07-05 & \texttt{2407.04258} & Reinforcement Learning for Unsupervised Video Summarization with Reward Generator Training \\
2021-09-03 & \texttt{2109.01309} & Unsupervised multi-latent space reinforcement learning framework for video summarization in ultrasound imaging \\
2021-06-19 & \texttt{2106.10528} & Video Summarization through Reinforcement Learning with a 3D Spatio-Temporal U-Net \\
2021-05-13 & \texttt{2105.06441} & DeepQAMVS: Query-Aware Hierarchical Pointer Networks for Multi-Video Summarization \\
2020-07-29 & \texttt{2007.14552} & Compare and Select: Video Summarization with Multi-Agent Reinforcement Learning \\
2020-05-19 & \texttt{2005.09531} & Ultrasound Video Summarization using Deep Reinforcement Learning \\
2020-01-12 & \texttt{2001.05864} & Weakly Supervised Video Summarization by Hierarchical Reinforcement Learning \\
2018-07-25 & \texttt{1807.09418} & Video Storytelling: Textual Summaries for Events \\
2018-07-09 & \texttt{1807.03089} & Video Summarisation by Classification with Deep Reinforcement Learning \\
2017-12-29 & \texttt{1801.00054} & Deep Reinforcement Learning for Unsupervised Video Summarization with Diversity-Representativeness Reward \\
\bottomrule
\end{tabular}
\end{tcolorbox}

\begin{tcolorbox}[
breakable,
title={Case 2: Claim-Comparison Query (\texttt{BQ\_000815})},
colback=gray!5,
colframe=gray!50!black,
fonttitle=\bfseries\small,
before skip=4pt,
after skip=6pt
]
\small
\textbf{Query.} Which papers report that transfer learning outperforms random initialization in model performance?\\
\textbf{Domain.} \texttt{cs.LG}; \textbf{Topic seed.} Transfer learning; \textbf{Constraint.} \texttt{comparison = outperforms random initialization}; \textbf{Gold answers.} 15.

\vspace{3pt}
\scriptsize
\setlength{\tabcolsep}{1pt}
\renewcommand{\arraystretch}{1.02}
\begin{tabular}{@{}p{0.16\linewidth}@{\hspace{1.5pt}}p{0.17\linewidth}@{\hspace{2pt}}p{0.62\linewidth}@{}}
\toprule
\textbf{Date} & \textbf{arXiv ID} & \textbf{Title} \\
\midrule
2025-11-06 & \texttt{2511.11622} & Small Vocabularies, Big Gains: Pretraining and Tokenization in Time Series Models \\
2024-10-10 & \texttt{2410.08194} & Features are fate: a theory of transfer learning in high-dimensional regression \\
2024-08-01 & \texttt{2408.00695} & Accelerating Full Waveform Inversion By Transfer Learning \\
2022-06-20 & \texttt{2206.09872} & A Neural Network Based Method with Transfer Learning for Genetic Data Analysis \\
2021-09-29 & \texttt{2109.14536} & PINNup: Robust neural network wavefield solutions using frequency upscaling and neuron splitting \\
2021-06-09 & \texttt{2106.04995} & Crosslingual Embeddings are Essential in UNMT for Distant Languages: An English to IndoAryan Case Study \\
2019-09-24 & \texttt{1909.10955} & Efficiently Reusing Old Models Across Languages via Transfer Learning \\
2019-08-26 & \texttt{1908.09883} & Transfer learning for scalability of neural-network quantum states \\
2018-11-03 & \texttt{1811.01137} & Transfer Learning in Multilingual Neural Machine Translation with Dynamic Vocabulary \\
2018-10-15 & \texttt{1810.06282} & Feature Representation Analysis of Deep Convolutional Neural Network using Two-stage Feature Transfer -An Application for Diffuse Lung Disease Classification- \\
2018-09-02 & \texttt{1809.00357} & Trivial Transfer Learning for Low-Resource Neural Machine Translation \\
2018-02-05 & \texttt{1802.01483} & Explicit Inductive Bias for Transfer Learning with Convolutional Networks \\
2017-10-11 & \texttt{1710.05726} & Convolutional Neural Networks for Histopathology Image Classification: Training vs. Using Pre-Trained Networks \\
2017-06-02 & \texttt{1706.00712} & Convolutional Neural Networks for Medical Image Analysis: Full Training or Fine Tuning? \\
2016-04-08 & \texttt{1604.02201} & Transfer Learning for Low-Resource Neural Machine Translation \\
\bottomrule
\end{tabular}
\end{tcolorbox}

\begin{tcolorbox}[
breakable,
title={Case 3: Scope-Control Query (\texttt{BQ\_001592})},
colback=gray!5,
colframe=gray!50!black,
fonttitle=\bfseries\small,
before skip=4pt,
after skip=6pt
]
\small
\textbf{Query.} I am looking for papers on software maintenance tools that exclude static analysis tools from their scope.\\
\textbf{Domain.} \texttt{cs.SE}; \textbf{Topic seed.} Software maintenance tools; \textbf{Constraint.} \texttt{exclude\_filtering = static analysis tools}; \textbf{Gold answers.} 12.

\vspace{3pt}
\scriptsize
\setlength{\tabcolsep}{1pt}
\renewcommand{\arraystretch}{1.02}
\begin{tabular}{@{}p{0.16\linewidth}@{\hspace{1.5pt}}p{0.17\linewidth}@{\hspace{2pt}}p{0.62\linewidth}@{}}
\toprule
\textbf{Date} & \textbf{arXiv ID} & \textbf{Title} \\
\midrule
2025-06-30 & \texttt{2506.24015} & Hierarchical Knowledge Injection for Improving LLM-based Program Repair \\
2025-06-16 & \texttt{2506.13182} & From Empirical Evaluation to Context-Aware Enhancement: Repairing Regression Errors with LLMs \\
2024-10-18 & \texttt{2410.14393} & Debug Smarter, Not Harder: AI Agents for Error Resolution in Computational Notebooks \\
2024-04-08 & \texttt{2404.05520} & The Fact Selection Problem in LLM-Based Program Repair \\
2023-04-24 & \texttt{2304.12015} & ITER: Iterative Neural Repair for Multi-Location Patches \\
2023-02-02 & \texttt{2302.01215} & Fixing Hardware Security Bugs with Large Language Models \\
2022-02-22 & \texttt{2202.10868} & Neural Program Repair: Systems, Challenges and Solutions \\
2021-12-03 & \texttt{2112.02125} & Examining Zero-Shot Vulnerability Repair with Large Language Models \\
2021-04-16 & \texttt{2104.08308} & Neural Transfer Learning for Repairing Security Vulnerabilities in C Code \\
2020-08-04 & \texttt{2008.01676} & Anchor: Locating Android Framework-specific Crashing Faults \\
2020-02-10 & \texttt{2002.03968} & E-APR: Mapping the Effectiveness of Automated Program Repair \\
2018-05-18 & \texttt{1805.07475} & Learning to Repair Software Vulnerabilities with Generative Adversarial Networks \\
\bottomrule
\end{tabular}
\end{tcolorbox}

\begin{tcolorbox}[
breakable,
title={Case 4: Setting-Anchored Query (\texttt{BQ\_001790})},
colback=gray!5,
colframe=gray!50!black,
fonttitle=\bfseries\small,
before skip=4pt,
after skip=6pt
]
\small
\textbf{Query.} Which papers study or evaluate privacy and security issues, policies, or privacy-preserving techniques in the healthcare domain?\\
\textbf{Domain.} \texttt{cs.CY}; \textbf{Topic seed.} Privacy policies; \textbf{Constraint.} \texttt{application\_domain = healthcare}; \textbf{Gold answers.} 19.

\vspace{3pt}
\scriptsize
\setlength{\tabcolsep}{1pt}
\renewcommand{\arraystretch}{1.02}
\begin{tabular}{@{}p{0.16\linewidth}@{\hspace{1.5pt}}p{0.17\linewidth}@{\hspace{2pt}}p{0.62\linewidth}@{}}
\toprule
\textbf{Date} & \textbf{arXiv ID} & \textbf{Title} \\
\midrule
2025-11-14 & \texttt{2511.11347} & Privacy Challenges and Solutions in Retrieval-Augmented Generation-Enhanced LLMs for Healthcare Chatbots: A Review of Applications, Risks, and Future Directions \\
2025-09-18 & \texttt{2509.14581} & Can I Trust This Chatbot? Assessing User Privacy in AI-Healthcare Chatbot Applications \\
2025-02-04 & \texttt{2502.02749} & Unveiling Privacy and Security Gaps in Female Health Apps \\
2024-10-18 & \texttt{2410.14607} & Evaluating Privacy Measures in Healthcare Apps Predominantly Used by Older Adults \\
2024-05-28 & \texttt{2405.17971} & A Qualitative Analysis Framework for mHealth Privacy Practices \\
2023-12-15 & \texttt{2312.10214} & Healthcare Policy Compliance: A Blockchain Smart Contract-Based Approach \\
2023-11-09 & \texttt{2311.05404} & A Survey on Privacy of Health Data Lifecycle: A Taxonomy, Review, and Future Directions \\
2023-07-24 & \texttt{2307.12542} & Client-Level Differential Privacy via Adaptive Intermediary in Federated Medical Imaging \\
2023-06-20 & \texttt{2306.11557} & A Comparative Audit of Privacy Policies from Healthcare Organizations in USA, UK and India \\
2023-06-10 & \texttt{2306.06448} & HIPAAChecker: The Comprehensive Solution for HIPAA Compliance in Android mHealth Apps \\
2023-06-09 & \texttt{2306.06033} & SoK: Analysis of User-Centered Studies Focusing on Healthcare Privacy \& Security \\
2023-02-08 & \texttt{2302.04208} & Exploratory Analysis of Federated Learning Methods with Differential Privacy on MIMIC-III \\
2022-11-21 & \texttt{2211.11434} & Privacy in Practice: Private COVID-19 Detection in X-Ray Images (Extended Version) \\
2022-05-06 & \texttt{2205.03168} & Defending against Reconstruction Attacks through Differentially Private Federated Learning for Classification of Heterogeneous Chest X-Ray Data \\
2021-06-26 & \texttt{2106.13973} & Benchmarking Differential Privacy and Federated Learning for BERT Models \\
2021-06-11 & \texttt{2106.06136} & Security and Privacy for Healthcare Blockchains \\
2020-09-17 & \texttt{2009.08294} & Robust Aggregation for Adaptive Privacy Preserving Federated Learning in Healthcare \\
2020-08-13 & \texttt{2008.05864} & An Empirical Evaluation of GDPR Compliance Violations in Android mHealth Apps \\
2019-10-02 & \texttt{1910.00962} & Privacy-preserving Federated Brain Tumour Segmentation \\
\bottomrule
\end{tabular}
\end{tcolorbox}

~\\


\section{Prompt Templates}
\label{app:prompts}

To make the LLM-assisted components of ScholarQuest transparent, we provide the main prompt templates used in query construction, answer discovery, agentic search, and relevance filtering. The templates serve the following roles:
\begin{itemize}[leftmargin=*]
    \item \textbf{Query Generation Prompt.} Generates one taxonomy-guided paper-retrieval query for each intent type from a given topic seed.
    \item \textbf{Query Rewrite Prompt.} Rewrites a benchmark query into multiple first-round search anchors while preserving the original retrieval intent.
    \item \textbf{Scorer Prompt.} Asks a relevance judge to assess whether a candidate paper satisfies the user query based on its title and abstract.
    \item \textbf{Strict Batch Filter Prompt.} Performs conservative batched relevance filtering for final answer curation, requiring explicit satisfaction of all core query constraints.
\end{itemize}

\begin{tcblisting}{
breakable,
title=Query Generation Prompt,
colback=gray!5,
colframe=gray!50!black,
before skip=1pt,
after skip=1pt,
listing only,
listing options={
basicstyle=\ttfamily\footnotesize,
breaklines=true,
columns=fullflexible,
keepspaces=true
}
}
You generate paper retrieval queries from one topic seed.

Return strict JSON only. The response must be one valid JSON object with this schema:
{
  "seed_id": "PTS_000001",
  "topic_seed": "large language model agents",
  "domain": "cs.AI",
  "queries": [
    {
      "category": "method_capability",
      "constraint_kind": "training_paradigm",
      "constraint_value": "reinforcement learning",
      "query": "Which papers study large language model agents trained with reinforcement learning?",
      "rationale": "reinforcement learning is a searchable technical constraint",
      "risk_flags": []
    }
  ]
}

Generate exactly one query for each query category:
- method_capability
- setting_anchor
- claim_comparison
- scope_control

Rules:
- Produce exactly four query objects, one per category.
- Each query must be a paper retrieval request.
- Keep each query concise, natural, and one sentence.
- Include one strong executable retrieval constraint per query.
- Stay faithful to the topic seed, domain, and ACM id.
- Avoid temporal wording such as recent, latest, since 2020, after 2020, or similar time constraints.
- Do not ask for analysis, advice, or long-form synthesis.
- Do not use fragment wording such as "Papers on ...".
- Vary the surface form across the four queries when natural.
\end{tcblisting}

\begin{tcblisting}{
breakable,
title=Query Rewrite Prompt,
colback=gray!5,
colframe=gray!50!black,
before skip=1pt,
after skip=1pt,
listing only,
listing options={
basicstyle=\ttfamily\footnotesize,
breaklines=true,
columns=fullflexible,
keepspaces=true
}
}
You rewrite one paper-search query into first-round search anchors.

Requirements:
- Return exactly 10 English queries.
- Keep every query close to the original intent.
- Vary the semantic angle, retrieval scope, terminology, method, setting, task, evidence type, or comparison focus.
- Make the 10 queries suitable as independent first-round search inputs for finding answer papers.
- Keep each query concise and retrieval-oriented.
- Avoid duplicates and near-duplicates.
- Avoid drifting into a different research problem.
- Do not mention citations, references, hops, or tool usage.
- Do not output markdown or explanations.

Return strict JSON only:
{"queries": ["...", "..."]}
\end{tcblisting}

\begin{tcblisting}{
breakable,
title=Scorer Prompt,
colback=gray!5,
colframe=gray!50!black,
before skip=1pt,
after skip=1pt,
listing only,
listing options={
basicstyle=\ttfamily\footnotesize,
breaklines=true,
columns=fullflexible,
keepspaces=true
}
}
You are an elite researcher in the field of AI, conducting research on {user_query}.
Evaluate whether the following paper fully satisfies the detailed requirements of the user query and provide your reasoning.
Ensure that your decision and reasoning are consistent.

Searched Paper:
Title: {title}
Abstract: {abstract}

User Query: {user_query}

Output format:+++++++++++++++++
Decision: True/False
Reason: ...
Decision:
\end{tcblisting}

\begin{tcblisting}{
breakable,
title=Strict Batch Filter Prompt,
colback=gray!5,
colframe=gray!50!black,
before skip=1pt,
after skip=1pt,
listing only,
listing options={
basicstyle=\ttfamily\footnotesize,
breaklines=true,
columns=fullflexible,
keepspaces=true
}
}
You are a strict relevance judge for academic paper filtering.

Your task is to evaluate whether each candidate paper strictly matches the same user query.
Judge based on the user query, paper title, and paper abstract.

Scoring rules:
- 2 = Strict match. The paper fully matches every core requirement in the query, including the target topic, method/task, setting, constraints, and any specified conditions. All important points in the query must be clearly supported by the title or abstract.
- 1 = Partial match or missing/violated constraint. The paper is related to part of the query, but at least one core requirement is missing, unclear, too broad, or contradicted. If a query constraint is not explicitly matched, assign 1 rather than 2.
- 0 = Mismatch. The paper does not match the query, or the main subject/intent differs from the query.

Confidence:
- high = The title and abstract provide enough evidence for the decision.
- medium = The evidence is somewhat incomplete but the decision is still reasonably supported.
- low = The decision is uncertain due to limited or ambiguous information.

Be conservative. Assign 2 only when the match is explicit and complete. Do not infer missing constraints from vague similarity.

Output valid JSON only:
{
  "results": [
    {
      "paper_index": 1,
      "arxiv_id": "...",
      "reason": "brief English reason based on the query, title, and abstract",
      "strict_score": 0,
      "confidence": "low"
    }
  ]
}

Return exactly one result for each input paper. Preserve the input paper_index and arxiv_id.
Within each result, place "reason" before "strict_score", and place "confidence" last.
Do not output markdown or extra text.

USER QUERY:
{query}

PAPER CANDIDATES JSON:
{papers_json}
\end{tcblisting}

\end{document}